# Demisability and survivability sensitivity to design-for-demise techniques

Mirko Trisolini[a], Hugh G. Lewis[b], Camilla Colombo[c]

[a]University of Southampton, University Road, SO17 1BJ, Southampton, United Kingdom.
m.trisolini@soton.ac.uk
[b]University of Southampton, University Road, SO17 1BJ, Southampton, United Kingdom,
H.G.Lewis@soton.ac.uk
[c]Politecnico di Milano, Via La Masa 34, 20156, Milano, Italy, camilla.colombo@polimi.it

**Abstract**

The paper is concerned with examining the effects that design-for-demise solutions can have not only on the demisability of components, but also on their survivability that is their capability to withstand impacts from space debris. First two models are introduced. A demisability model to predict the behaviour of spacecraft components during the atmospheric re-entry and a survivability model to assess the vulnerability of spacecraft structures against space debris impacts. Two indices that evaluate the level of demisability and survivability are also proposed. The two models are then used to study the sensitivity of the demisability and of the survivability indices as a function of typical design-for-demise options. The demisability and the survivability can in fact be influenced by the same design parameters in a competing fashion that is while the demisability is improved, the survivability is worsened and vice versa. The analysis shows how the design-for-demise solutions influence the demisability and the survivability independently. In addition, the effect that a solution has simultaneously on the two criteria is assessed. Results shows which, among the design-for-demise parameters mostly influence the demisability and the survivability. For such design parameters maps are presented, describing their influence on the demisability and survivability indices. These maps represent a useful tool to quickly assess the level of demisability and survivability that can be expected from a component, when specific design parameters are changed.

## 1 Introduction

During the past two decades, the attention towards a sustainable exploitation of the space environment has raised steadily. The space around the Earth and beyond has been the theatre of remarkable achievements in the past sixty years but has also suffered from the consequences of the thousands of missions that have flown since then. Decommissioned satellites, spent upper stages, other mission related objects, and fragments generated by collisions and explosions of spacecraft and upper stages pollute the space environment in the form of space debris. Space debris is recognised as a major risk to space missions, in fact, an object of just 1 cm in size can cause the disruption of a satellite, and smaller particles can still have enough energy to produce failures on components critical to the mission success. Recent studies about the evolution of the space environment have shown a continuum increase in the population of space debris [1-4], and the amount of debris is expected to keep growing unless mitigation measures are implemented in the following years. The most effective

---

Abbreviations

DRAMA: Debris Risk Assessment and Mitigation Analysis
MASTER: Meteoroid and Space Debris Terrestrial Environment Reference
MIDAS: MASTER-based Impact Flux and Damage Assessment Software
LMF: Liquid Mass Fraction
PNP: Probability of no-penetration

among these mitigation measures is the limitation of the long-term presence of spacecraft and upper stages in the Low Earth Orbit (LEO) and Geostationary Orbit (GEO) protected regions [5]. This in turn means that a spacecraft has to be removed from its operational orbit after its decommissioning, either by placing it in a graveyard orbit or by allowing it to re-enter into the Earth's atmosphere. For LEO spacecraft, the preferred scenario is to design a disposal by re-entry within 25 years from its decommissioning in order for the mission to comply with the 25-year rule [6]. However, when a spacecraft is to be disposed through re-entry it has also to satisfy the requirement for the limitation of the risk of human casualty on the ground associated to the debris surviving the re-entry. This can be either achieved performing a controlled re-entry, where the spacecraft is guided to impact in the ocean or not populated areas, or through an uncontrolled re-entry, where the vehicle is left to re-enter without any guidance. In the latter case, the surviving mass of the spacecraft has to be low enough to comply with the regulation on the casualty risk expectation that has to be below the threshold of $10^{-4}$. Controlled re-entries have a larger impact on the mission performance with respect to uncontrolled ones, as they require a larger amount of fuel to be performed and a higher level of reliability. The spacecraft, in fact, has to carry enough fuel to perform the final disposal manoeuvre at the end of its operational life. On average between 10% to 40% of the spacecraft initial mass survives re-entry [7]. As a consequence, in order to exploit the advantages of an uncontrolled re-entry strategy in terms of its simplicity and its cost (necessity of new AOCS modes and the possibility to move to a bigger launcher) [7, 8] and still meet the casualty risk constraint, design solutions that favours the demisability of the spacecraft and its components can be adopted. This approach is known as design-for-demise, which is the procedure to consider, since the early stages of the mission planning, design options that will allow the demise of the spacecraft in the atmosphere. Among the specific methods employed in designing spacecraft parts to demise, the following can be identified [7, 9-11]: selection of the material, use of multiple materials, optimisation of the shape, size, thickness of the component, and changing the component location.

The attention towards design-for-demise has been increasing in the past few years with a growing effort to find solutions to increase the demisability of spacecraft parts and structures. In particular, the European Space Agency, through the Clean Space initiative [12-14] is investing into new demisable solutions for particularly sensitive components such as tanks and reaction wheels. Nonetheless, spacecraft and components designed to demise, still has to survive to the large amount of space debris and micrometeoroids that can penetrate the spacecraft structure and damage components and subsystems. Ensuring the spacecraft reliability against space debris impacts during its operational life is important. It is also necessary to adequately protect the spacecraft after its operational life. Although in this case the main mission of the spacecraft is concluded, it still has to carry out a disposal strategy and comply with the regulations for space debris limitation. In fact, the most critical components inside the satellite still have to be protected in order to avoid possible debris impact induced explosions or break-ups (especially for sensitive components such as tanks and batteries) or compromising the end-of-life disposal strategy. The study here presented focuses on the effect of space debris on LEO spacecraft, neglecting the effects of micrometeoroids, which have lower densities and energies at the altitudes considered in the study.

Design-for-demise solutions can be used to modify the characteristics of spacecraft components; as such, they can also influence the components survivability against the impact from space debris. For instance, can changing the material of a tank to make it demisable compromise its resistance to debris impacts? Can the design of a more demisable configuration increase the vulnerability of the spacecraft to the debris environment?

Considering that the design-for-demise is a relatively new field of study, the aim of this paper is to analyse how a design-for-demise oriented approach can influence other subsystems and other mission requirements. Furthermore, considering that the design requirements connected to the demisability and the ones connected to the survivability (the ability of an object to withstand debris impacts) appears to be conflicting in nature, it is interesting to investigate in which way they can mutually influence each other and to how and to what extent they are influenced by common design choices.

The paper presents a re-entry model used to compute the demisability of a spacecraft its disposal through atmospheric re-entry. In addition, a model developed to assess the survivability of spacecraft components against the impact from space debris during its operational life is presented. A demisability and a survivability indices have also been developed in order to quantify the level of demisability and survivability of a given spacecraft component.

The design parameters affecting both the demisability and the survivability of a component are first identified. The sensitivity of the demisability and the survivability index to the design choices is then analysed. Indeed, the specific re-entry conditions can influence the demisability of an object, in the same way as the operational orbit selection (altitude and inclination) and the mission lifetime can affect the survivability. Finally, the effect of the design-for-demise options on both the demisability and the survivability index is studied using a representative component. Using a spacecraft tank as the reference component, the influence of the design-for-demise options is studied varying each parameter in order to understand to what extent and how they influence the demisability and survivability indices.

## 2  Re-entry model

The developed re-entry model can be classified under the category of object-oriented codes. It is able to simulate the three degree-of-freedom trajectory for elementary geometrical shapes representative of spacecraft components, i.e., sphere, cylinder, flat plate, box, assuming a predefined random tumbling motion. The ablation is analysed with a lumped mass model; when the melting temperature is reached, the mass is considered to vary as a function of the heat of ablation of the material. All the material properties are temperature independent and have an average value from the ambient temperature up to the melting temperature. Average drag coefficients, shape factors, and correlations needed to describe the aerodynamic and aero-thermodynamic behaviour of the object were taken or derived from the literature. The demise is assessed as the ratio between the residual mass of the object after the re-entry and its initial mass.

The model is the result of a major work of unification of the different sources of information for the heat rate correlations, drag coefficients expressions, and material behaviour sparsely found in the literature. The retrieved information had also to be adapted to the application in exam as it is described in the following paragraphs.

### 2.1  Re-entry environment

During the descent trajectory, a satellite experiences the effects of the surrounding environment in the form of forces and moments acting on it and influencing its motion. The main sources of external forces are the pressure forces (lift and drag) due to the aerodynamic interaction between the satellite and the Earth's atmosphere, and the gravitational forces generated by the effect of the Earths' gravitational field on the spacecraft. A zonal harmonic gravity model up to degree 4 is adopted in the current version of the software. The radial and tangential acceleration components acting on the satellite due to gravity can be expressed as [15]

$$g_C = \frac{\mu_e}{r} \left\{ 1 - \frac{3}{2} J_2 \left( \frac{R_e}{r} \right)^2 [3\cos^2\phi - 1] - 2 J_3 \left( \frac{R_e}{r} \right)^3 [5\cos^3\phi - 3\cos\phi] \right. \\ \left. - \frac{5}{8} J_4 \left( \frac{R_e}{r} \right)^4 [35\cos^4\phi - 30\cos^2\phi + 3] \right\} \quad (1)$$

$$g_\delta = -\frac{3\mu_e}{r^2} \left( \frac{R_e}{r} \right)^2 \sin\phi\cos\phi \left\{ J_2 + \frac{1}{2} J_3 \left( \frac{R_e}{r} \right) [5\cos^2\phi - 1] \right. \\ \left. + \frac{5}{6} J_4 \left( \frac{R_e}{r} \right)^2 [7\cos^2\phi - 1] \right\} \quad (2)$$

where $\mu_e$ is the gravitational parameter of the Earth, $R_e$ is the Earth's radius, $r$ is the distance between the centre of the Earth and the satellite, $\phi$ is the colatitude, and $J_k$ (k = 1,..,4) are the zonal harmonics coefficients, also known as Jeffery constants.

The atmospheric model implemented in the software is based on the 1976 U.S. Standard Atmosphere [16]. The Earth's atmosphere is divided into two main zones: the lower atmosphere, which extends from the surface to a geometric altitude of 86 km, and the upper atmosphere, which ranges from 86 km up to 1000 km. Each of the two zones is further divided into layers. Within each layer, the temperature is represented with a predefined function of the altitude. Pressure and density are then derived accordingly as functions of the altitude.

The lower atmosphere is divided into seven layers. In each layer, the temperature is assumed to vary linearly with respect to the geopotential altitude. Knowing the temperature profile is possible to express the variation of the atmospheric density and pressure with the altitude.

For the upper atmosphere, a similar approach is followed [17]. Four temperature profiles are used to express the atmospheric temperature variation with the altitude. The value of the density and pressure with the altitude is instead obtained through a cubic interpolation of the tabulated data provided in the 1976 U.S. Standard Atmosphere. For such an interpolation, the upper atmosphere is divided into 22 layers and for each layer the base values of pressure and density and their derivatives are defined. These values are then used inside the interpolation function to compute the values of atmospheric pressure and density at a user specified altitude.

*2.2 Equations of motion*

The dynamic model is a three degree-of-freedom schematisation that describes the evolution of the trajectory considering the satellite as a point mass with a predefined attitude. As the aerodynamic forces on the spacecraft are due to the motion of the vehicle relative to the atmosphere of the planet, it is convenient to express the equations of motion in a reference frame rotating with the atmosphere. Because a planet's atmosphere rotates with it, it is possible to define a planet-fixed reference frame in order to express the equations of atmospheric flight. The kinematic equations of motion are as follows [15]:

$$\dot{r} = V_\infty \sin\gamma \tag{3}$$

$$\dot{\delta} = \frac{V_\infty}{r} \cos\gamma \cos\alpha \tag{4}$$

$$\dot{\lambda} = \frac{V_\infty \cos\gamma \sin\alpha}{r \cos\delta} \tag{5}$$

where $V_\infty$, $\gamma$, and $\alpha$ are the relative velocity magnitude, the flight path angle and the azimuth in the local horizon plane respectively, and $\delta$ is the longitude.

The dynamic equations of motion can then be written as follows [15]:

$$\dot{V}_\infty = -\frac{D}{m} - g_C \sin\gamma + g_\delta \cos\gamma \cos\alpha \\ - \omega^2 r \cos\delta (\cos\gamma \cos\alpha \sin\delta - \sin\gamma \cos\delta) \tag{6}$$

$$V_\infty \cos\gamma \, \dot{\alpha} = \frac{V_\infty}{r} \cos^2\gamma \sin\alpha \tan\delta - g_\delta \sin\alpha + \omega^2 r \sin\alpha \sin\delta \cos\delta \\ - 2\omega V_\infty (\sin\gamma \cos\alpha \cos\delta - \cos\gamma \sin\delta) \tag{7}$$

$$V_\infty \dot{\gamma} = \left(\frac{V_\infty^2}{r} - g_C\right) \cos\gamma + \frac{L}{m} - g_\delta \sin\gamma \cos\alpha \\ + \omega^2 r \cos\delta (\sin\gamma \cos\alpha \sin\delta + \cos\gamma \cos\delta) + 2\omega V_\infty \sin\alpha \cos\delta \tag{8}$$

where $\omega$ is the angular rotational velocity of the planet, $L$ is the lift force, $D$ is the drag force, and $g_c$ and $g_\delta$ are the component of the gravitational acceleration.

As only the uncontrolled re-entry of satellites is taken into account, the thrust component in the dynamic equations is neglected and only the aerodynamic and the gravitational forces are considered. In general, for an uncontrolled re-entry, also the lift can be assumed negligible as the satellites usually do not have aerodynamic shapes and tend to assume a random attitude during the re-entry. The complete set of governing Eqs. (3) - (8) can be integrated in time to obtain the position and the velocity of the spacecraft at every time instant along the trajectory.

*2.3  Aerodynamics and aerothermodynamics*

Aerodynamics and aerothermodynamics represents the core of the demisability analysis because allow the computation of the forces acting on the satellite influencing its trajectory, as well as the heat load the satellite experiences. As an object-oriented approach is followed, a series of motion and geometry averaged coefficients for both the aerodynamics and aerothermodynamics are necessary to describe the aerodynamics and the head loads acting on re-entering satellites. As the main phenomena influencing aerodynamic and thermal loads happen in the hypersonic regime, only hypersonic aerodynamics and aerothermodynamics are considered in the model. This is also a common choice in most of the destructive re-entry software [18].

*2.3.1  Aerodynamics*

Aerodynamic forces are usually expressed in terms of non-dimensional coefficients. The drag coefficient ($C_D$) is expressed as follows:

$$C_D = \frac{D}{q_\infty S} \qquad (9)$$

where $q_\infty = 1/2 \cdot \rho_\infty V_\infty^2$ is the dynamic pressure, $\rho_\infty$ and $V_\infty^2$ are the free-stream density and velocity respectively, and $S$ is the reference cross-sectional area of the re-entering object.

The developed software employs a set of drag coefficients for the four possible geometrical shapes available (i.e., sphere, cylinder, box, and flat plate). Both the free-molecular and continuum flow regimes are taken into account. Expressions for the drag coefficients for the continuum and free-molecular flux regimes have been developed or derived from the literature and bridging functions are utilised for the description of the aerodynamics in the transitional regime.

**Table 1: Free-molecular drag coefficients for sphere, cylinder, box, and flat plate.**

| Shape | Free-molecular $C_D$ | Reference Area |
|---|---|---|
| Sphere | 2.0 | $\pi \cdot R^2$ |
| Cylinder | $1.57 + 0.785 \cdot \dfrac{D}{L}$ | $2 \cdot R \cdot L$ |
| Box | $1.03 \cdot \left[ \dfrac{A_x + A_y + A_z}{A_y} \right]$ | $A_y$ * |
| Flat Plate | 1.03 | $A$ |

In Table 1 are summarised the free-molecular drag coefficients used in the code developed. The drag coefficient for spheres is taken from Masson, Morris and Bloxsom [19], where it is assumed to have a constant value. For cylinders [20], the drag coefficient is a function of the ratio between the cylinder diameter ($D$) and length ($L$). For flat plates, a constant value is adopted Hallman and Moody

---

* $A_y$ is the median surface area of the face of the box, such that. $A_x \geq A_y \geq A_z$.

[21]. For box shaped object, the $C_D$ is related to the drag coefficient of a flat plate perpendicular to the free stream by averaging it over the faces of the box.

Table 2: Continuum drag coefficients for sphere, cylinder, box, and flat plate.

| Shape | Continuum $C_D$ | Reference Area |
|---|---|---|
| Sphere | 0.92 | $\pi \cdot R^2$ |
| Cylinder | $0.7198 + 0.3260 \cdot \dfrac{D}{L}$ | $2 \cdot R \cdot L$ |
| Box | $0.46 \cdot \left[ \dfrac{A_x + A_y + A_z}{A_y} \right]$ | $A_y$ |
| Flat Plate | 0.46 | $A$ |

Table 2 shows the continuum drag coefficients for the same geometries. The relationships were derived from the same respective references used for the free-molecular drag coefficients. It is also necessary to provide relationships for the transition between the free-molecular and the continuum flow regimes. Two bridging functions are implemented as follows:

$$C_{D,t} = C_{D,c} + \left( C_{D,fm} - C_{D,c} \right) \cdot \left[ \sin \left( \pi \left( 0.5 + 0.25 \cdot \log_{10}(Kn) \right) \right) \right]^3 \tag{10}$$

$$C_{D,t} = C_{D,c} + \left( C_{D,fm} - C_{D,c} \right) \cdot \exp\left[ -\frac{\rho_{atm} \cdot r}{10 \cdot \rho_s \cdot \lambda_s} \right] \tag{11}$$

where $C_{D,fm}$ and $C_{D,c}$ are the free-molecular and the continuum drag coefficients respectively, and $Kn$ is the Knudsen number of the flow surrounding the re-entry object.

Eq. (11) [19] is the bridging function adopted for spheres, where $r$ is the radius of the sphere, $\rho_{atm}$ is the atmospheric density, $\rho_S$ is the air density after the normal shock, and $\lambda_S$ is the mean free path after the shock. Eq. (10) is the bridging function [22, 23] used for all the remaining shapes.

*2.3.2 Aerothermodynamics*

In order to analyse the demisability of re-entering objects the heat load acting on them during the descent trajectory needs to be computed. As the attitude of the object is predefined and assumed to be random tumbling, the heat load is computed using averaging factors, which take into account the shape and the motion of the object. The averaging factors, also referred as shape factors ($\overline{F}_q$), provide the relationship between a reference heat load ($\dot{q}_{ref}$) and the average heat rate on the object as a function of its geometrical shape and the attitude motion. The reference heat load depends on the flux regime: corresponds to the stagnation-point heat rate on a sphere ($\dot{q}_{ss}$) for continuum flows and to the heat rate on a flat plate perpendicular to the flux ($\dot{q}_{fp}^{\perp}$) for free-molecular flows. In general, the average heat rate on an object can be expressed as:

$$\dot{q}_{av} = \overline{F}_q \cdot \dot{q}_{ref} \tag{12}$$

*2.3.2.1 Free-molecular aerothermodynamics*

Following the approach described in Klett [20] the reference heat rate for free-molecular flows can be expressed as:

$$\dot{q}_{fp}^{\perp} = 11356.6 \times \left[ \frac{a \cdot \rho_{\infty} \cdot V_{\infty}^3}{1556} \right] \tag{13}$$

where $\rho_\infty$ and $V_\infty$ are the free-stream density and velocity respectively, and $a$ is the thermal accommodation coefficient whose value is assumed to be constant and equal to 0.9 [24].

Very limited data is available in the literature for shape factors in the free-molecular flow; therefore, some assumptions had to be made. For spheres in free-molecular flow, as no data could be found in the literature, the free-molecular shape factor for a random tumbling disk [20] was adopted. The corresponding value is

$$\overline{F}^{s}_{q,fm} = 0.255 \tag{14}$$

For cylinders the method presented in Klett [20] is used to compute the free-molecular shape factor, leading to the following equation:

$$\overline{F}^{c}_{q,fm} = \frac{\left(2 \cdot \overline{F}^{end}_{q,fm} \cdot A_{end} + \overline{F}^{side}_{q,fm} \cdot A_{side}\right)}{A_{tot}} \tag{15}$$

where $A_{end} = \pi R^2$ and $A_{side} = 2\pi RL$ are the end and side areas of the cylinder, and $A_{tot} = 2A_{end} + A_{side}$ is the total area of the cylinder. $\overline{F}^{end}_{q,fm}$ and $\overline{F}^{side}_{q,fm}$ are the shape coefficients relative to the end and side of the random tumbling cylinder respectively, and can be expressed as:

$$\overline{F}^{end}_{q,fm} = 0.255 \tag{16}$$

$$\overline{F}^{side}_{q,fm} = 0.785 \cdot Y + 0.500 \cdot Z \tag{17}$$

where $Y$ and $Z$ can be derived from Klett [20]. $Y$ is a curve representing the free-molecular flow ratio of the average heating on the sides of a rotating side-on cylinder to the heating on surfaces perpendicular to the flow. $Z$ describes instead the free-molecular flow ratio of the heating on surfaces parallel to the flow to the heating on surfaces perpendicular to the flow

For box shaped structure, the procedure outlined in Hallman and Moody [21] is adopted, where the box is reduced to an equivalent cylinder and Eq. (15) can still be used. The equivalence between the box and the cylinder is accomplished in the following way: assuming a box with side lengths equal to $l$, $w$, $h$ and following the relationship $l > w > h$, the equivalent length, and the equivalent radius of the cylinder are respectively $l_e = l$ and $r_e = \sqrt{w \cdot h}$.

Finally, for flat plates the constant value derived for random tumbling disks [20] is adopted.

$$\overline{F}^{fp}_{q,fm} = 0.255 \tag{18}$$

Eq. (18) is actually obtained following the theory of Oppenheim [20], which was originally developed for flat plates, and extended by Klett to the disks. Consequently, the extension of such value to flat plates is justified.

*2.3.2.2 Continuum flow aerothermodynamics*

As introduced before, in continuum flux the shape factors are computed taking as reference the heat rate at the stagnation point of a sphere. A large variety of correlations is available in the literature [25, 26]. The developed software adopts the Detra, Kemp Riddel (DKR) [25] for continuum flow regimes:

$$\dot{q}_{ss} = 1.99876 \times 10^8 \sqrt{\frac{0.3048}{r_n}} \sqrt{\frac{\rho_\infty}{\rho_{SL}}} \cdot \left(\frac{V_\infty}{7924.8}\right)^{3.15} \frac{h_s - h_w}{h_s - h_{w_{300}}} \tag{19}$$

The continuum shape factor for a random tumbling sphere was derived from an average of shape coefficients provided in the literature [27, 28], obtaining a value of:

$$\overline{F}_{q,c}^{s} = 0.234 \tag{20}$$

For cylindrical shape, the procedure followed in Klett [20] is adopted, obtaining a shape factor as:

$$\overline{F}_{q,fm}^{c} = \frac{\left(2 \cdot \overline{F}_{q,c}^{end} \cdot A_{end} + \overline{F}_{q,c}^{side} \cdot A_{side}\right)}{A_{tot}} \tag{21}$$

which is equivalent to the free molecular case, except for the coefficients $\overline{F}_{q,c}^{end}$ and $\overline{F}_{q,c}^{end}$ that represent the shape factors for the end and the side of a random tumbling cylinder in continuum flow as follows:

$$\overline{F}_{q,c}^{side} = 0.179 + 0.333 \cdot B \tag{22}$$

$$\overline{F}_{q,c}^{end} = 0.323 \tag{23}$$

where $B$ [20] is a curve representing the ratio of the average heating on the side of an end-on cylinder to the stagnation point heating on a sphere of the same radius.

For box shaped structure, an equivalent cylinder is defined in the same way as before and the Eq. (21) is used to compute the shape factor.

For the flat plate, as no specific results were found in the literature, the shape coefficient for a random tumbling disk was taken as reference. An equivalent disk is defined to compute the shape coefficient for a flat plate. For the heat load on a flat plate, we have:

$$\dot{Q}_{fp} = \overline{F}_{q,c}^{fp} A_{w}^{fp} \dot{q}_{SS} \tag{24}$$

where $A_{w}^{fp} = 2 \cdot (l \cdot w + l \cdot t + w \cdot t)$ is the wet area of the entire flat plate. To build the equivalent disk we assume an equivalent radius equal to $r_e = w/2$. For a disk the heat load would be:

$$\dot{Q}_{fp} = \overline{F}_{q,c}^{d,eq} A_{w}^{d,eq} \dot{q}_{SS} \tag{25}$$

As we are using an equivalent disk, we want it to have the same heat load of the flat plate. It is thus possible to equate Eq. (24) and Eq. (25) in order to obtain the shape coefficient for a flat plate as follows:

$$\overline{F}_{q,c}^{fp} = \overline{F}_{q,c}^{d,eq} \cdot \frac{A_{w}^{d,eq}}{A_{w}^{fp}} \tag{26}$$

where $\overline{F}_{q,c}^{d,eq}$ is the shape coefficient of the equivalent disk that, for a random tumbling disk, is independent on the radius and can be obtained from Klett [20] where a value of 0.323 is reported.

In the same way as for the drag coefficients, a bridging function has to be used to obtain the shape factor in the transition regime. The bridging function adopted in the code has the following expression [29]:

$$\dot{q}_t = \dot{q}_c \Bigg/ \sqrt{1 + \left(\frac{\dot{q}_c}{\dot{q}_{fm}}\right)^2} \tag{27}$$

*2.4 Ablation model*

To describe the mass loss during the demisability analysis of a re-entering object is a lumped mass model is used. In this schematisation, the object is considered to have a certain mass with a uniform temperature. As the heat load increases during the atmospheric descent, the temperature of the object increases until it reaches the melting temperature of the material. The temperature variation during the heating phase can be described with the equation:

$$\frac{dT_w}{dt} = \frac{A_w}{m(t) C_{p,m}} \left[ \dot{q}_{av} - \varepsilon \sigma T_w^4 \right] \quad (28)$$

where $T_w$ is the temperature of the object at a certain time instant, considered uniform for the entire volume. $A_w$ is the wetted area, $m$ is the instantaneous mass of the object, $C_{p,m}$ is the specific heat capacity of the material, $\varepsilon$ is the emissivity of the material, $\sigma$ is the Stefan-Boltzmann constant, and $\dot{q}_{av}$ is the heat flux on the object with the shape and attitude dependant averaging factor already taken into account. The model takes into account the variation with time of the mass of the object and the heat loss due to the re-radiation, but it does not take into account oxidation heating.

Once the melting temperature is reached, the object starts melting and loosing mass at a rate that is proportional to the net heat flux on the object and to the heat of fusion ($h_m$) of the material as follows:

$$\frac{dm}{dt} = -\frac{A_w}{h_m} \left[ \dot{q}_{av} - \varepsilon \sigma T_w^4 \right] \quad (29)$$

It is important to highlight that this model assumes that the conduction inside the object is infinite so that the temperature is uniform everywhere in the volume of the object. This is a good approximation for metallic structure. For non-metallic materials, such as composites, this kind of approach is adapted using an equivalent metal approach and defining equivalent properties for the material under consideration [30, 31].

*2.5 Material database*

The material database used for the demise computation is derived from the NASA Debris Assessment Software 2.0 (DAS 2.0) [24, 32]. The material properties required for the demisability analysis are density, heat capacity, melting temperature, heat of fusion, and emissivity. All the material properties are assumed temperature independent. When material properties such as the emissivity where not available in the DAS database, they were taken from the MatWeb database [33]. The properties for the set of materials used in the present work is summarised in Table A1.

*2.6 Model validation*

In this paragraph a comparison is presented between the presented model and the results obtained by the re-entry code Spacecraft Aerothermal Model (SAM) [28] from Belstead Research [34] for a variety of standard test cases presented during the First Demise Workshop [35]. All the simulation have been performed with the initial conditions summarised in Table 3, for four different geometrical shapes (sphere, cylinder, box and flat plate) and for two materials (aluminium 7075-T6 and titanium 6Al-4V). The characteristics of the geometrical shapes considered are summarised in Table 4 together with the attitude motion assumed during the re-entry.

**Table 3: Initial conditions for re-entry comparisons.**

| Initial Conditions | |
|---|---|
| Longitude (deg) | 0 |
| Latitude (deg) | 0 |
| Altitude (km) | 120 |
| Velocity (m/s) | 7273 |
| Heading (deg) | 42.35 |
| Flight Path Angle (deg) | -2.612 |

The results obtained from the comparison are showed in Fig. 1 and Fig. 2 for the variation of the temperature profile with the altitude during the re-entry. The four basic geometrical shapes described in Table 4 are analysed for both the aluminium and titanium case and compared against the results of the software SAM.

Table 4: Geometry analysed.

| Shape | Dimensions (m) | Motion |
|---|---|---|
| Sphere | $R = 0.5$ x $t = 0.03$ | Random Tumbling |
| Cylinder | $R = 0.5$ x $L = 1.0$ x $t = 0.03$ | Random Tumbling |
| Box | $L = 1.0$ x $W = 1.0$ x $H = 1.0$ | Random Tumbling |
| Flat Plate | $L = 1.0$ x $W = 1.0$ x $t = 0.03$ | Random Tumbling |

Fig. 1 and Fig. 2 show that the model performs well against the software SAM and the differences observable in the temperature profiles are all inside the usual level of uncertainty of the majority of the destructive oriented codes. In fact, the differences between the stagnation point correlations are of the order of 8% [36] and the variability among the shape factor can be up to 20% [27]. The data used for the comparison are publicly available from the Belstead Research website [34].

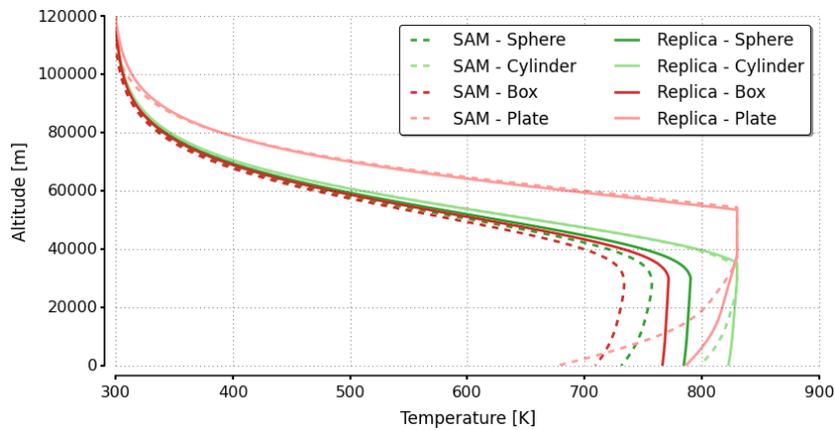

Fig. 1: Comparison of altitude temperature profiles for aluminium objects of different shapes.

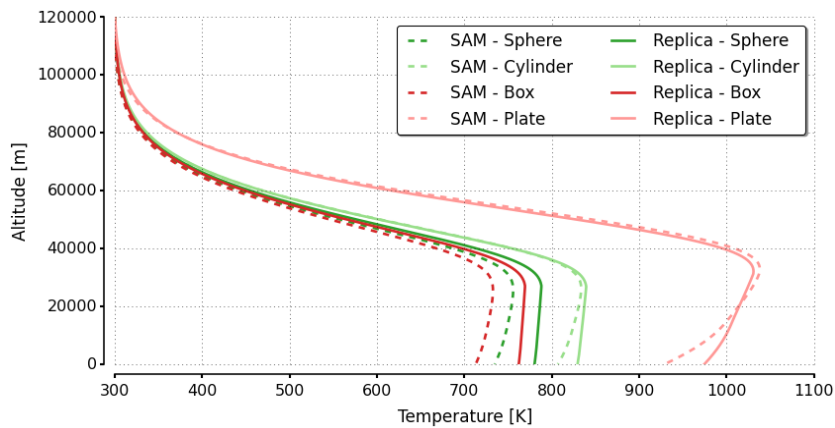

**Fig. 2: Comparison of altitude temperature profiles for aluminium objects of different shapes.**

*2.7  Demise criterion*

The previously described object oriented destructive re-entry code has been developed in order to be able to assess the demisability of spacecraft components and structures as a function of the characteristics of the objects such as the material, the geometry and the initial conditions at the re-entry. As a measure of the demisability of a component, the fraction of the mass that demises during the re-entry is taken into account, and is referred to as the Liquid Mass Fraction (LMF).

$$LMF = 1 - \frac{m_f}{m_{in}} \tag{30}$$

A value of 1.0 of the index corresponds to complete demise, whereas a value of 0 to complete survival. It is also taken into account the 15 J limit in the index so that when the re-entering object has an impact energy lower than 15 J the index is considered to be equal to 1.0.

# 3  Survivability model

The second model developed is a survivability model, which assess the vulnerability of a spacecraft against the impact from space debris. As this model is used in combination with the demisability model, a driving requirement to its development is to maintain characteristics similar to the demisability model, for what concerns both its complexity and computational time. Software such ESABASE/DEBRIS [37] and NASA BUMPER [38] are already available, however they are complex and they use computationally expensive methodology such as ray tracing methods [39]. The proposed approach, on the other hand, is tailored to be compatible with the object-oriented approach used for the demisability analysis, adopting a new strategy to compute the effects of the debris fluxes on a configuration. The computational model adopted starts with a panelised representation of the spacecraft structure, where to each panel are assigned the geometrical and material properties needed for the impact analysis. The geometrical shapes available are the same available in the demisability model (i.e. sphere, box, cylinder, flat plate). The core of the model is then the representation of the debris environment using *vector flux elements*, which describe the debris flux coming from a space sector in terms of flux, velocity, and direction. Finally, the impact of the debris onto the structure is evaluated using ballistic limit equations (BLE) [40-43], and the survivability is computed determining the probability of no-penetration (PNP) on the structure

*3.1  Survivability environment*

The space environment around the Earth is populated by a large amount of man-made debris and is also characterised by the presence of natural particles such as meteoroids. It is thus necessary to have a description as realistic as possible of the natural and man-made particulate environment around the Earth. The ESA software MASTER-2009 [44] provides the description of the debris environment via flux predictions on user defined target orbits. Both 2D and 3D flux distributions are available in MASTER-2009, as a function of many parameters such as the particle diameter, the impact velocity, etc. The specific distribution needed for the survivability computation [37] are the flux vs particle diameter, flux vs impact elevation, flux vs impact elevation vs impact azimuth, and the flux vs impact velocity vs impact azimuth. The user has to provide MASTER-2009 with the ranges and the subdivisions for the specified distributions. For the analysis presented the ranges and subdivisions are summarised in Table 5.

**Table 5: Ranges and subdivisions adopted in the paper.**

| Distribution | Range | Subdivisions |
|---|---|---|
| Flux vs Particle Diameter | $[10^{-4}, 0.1]$ | 50 |

| Flux vs Impact Elevation | [-90°, 90°] | Every 5° |
| Flux vs Impact Azimuth | [-180°, 180°] | Every 5° |
| Flux vs Impact Velocity | 0 - 40 km/s | Every 1 km/s |
| Flux vs Impact Elevation vs Impact Azimuth | Same as 2D distribution | Same as 2D distribution |
| Flux vs Impact Azimuth vs Impact Velocity | Same as 2D distribution | Same as 2D distribution |

*3.2  Ballistic limit equations*

Ballistic limit equations are relationships used to assess the damage caused by debris impacts on spacecraft structures as a function of the characteristics of the impact (velocity and direction) and of the target (material, dimensions, thickness, and shielding option). The output of a ballistic limit equation is the critical diameter, which is the diameter of the particle above which the structure suffers a damage. Possible damage modalities are penetration, detached spall, and incipient spall. The different modalities can be selected as a function of the particular component considered, whether it is a pressure vessel, a battery assembly or some other kind of component.

The BLEs for single wall structures adopted in the developed software are presented for completeness in the following. They are the reference equations for NASA's ballistic limit and debris impact analysis [40] [43]. Different equations are used for structures of different materials. For single wall aluminium structures, the BLE corresponds to

$$d_c = \left[ \frac{t_s}{k} \frac{HB^{1/4} (\rho_s / \rho_p)^\alpha}{5.24 \cdot (V_p \cos\theta / C)^{2/3}} \right]^{18/19} \tag{31}$$

with *HB* Brinell hardness of the material, $\rho_S$ and $\rho_p$ density of the shield and of the projectile respectively expressed in g/cm$^3$, $V_p$ particle relative impact velocity in km/s, *C* speed of sound in the material considered in km/s, $t_S$ thickness of the shield in cm, and $\theta$ impact angle. The value of the exponent $\alpha$ depends on the ratio between the particle density and the shield density and the constant *k* depends on the type of failure mode considered: $k = 3.0$ for incipient spall, $k = 2.2$ for detached spall, and $k = 1.8$ for perforation. Analogous equations can be derived for titanium (32), stainless steel (33), and CFRP (34) single wall structures as follows:

$$d_c = \frac{t_s}{k} \frac{HB^{1/4} (\rho_s / \rho_p)^{1/2}}{5.24 \cdot (V_p \cos\theta / C)^{2/3}} \tag{32}$$

$$d_c = \left[ \frac{t_s}{0.621} \frac{(\rho_s / \rho_p)^{1/2}}{(V_p \cos\theta / C)^{2/3}} \right]^{18/19} \tag{33}$$

$$d_c = \frac{t_s}{k} \frac{(\rho_s / \rho_p)^{1/2}}{K_{CFRP} (V_p \cos\theta)^{2/3}} \tag{34}$$

In Eq. (33) for the stainless steel, the hardness of the material is included in the coefficient. In addition, no distinction is made between possible failure modes, as only perforation damage is considered. Eq. (34) was derived [43] for CFRP introducing the parameter $K_{CFRP}$ that takes into account the effects of the material properties such as the hardness, the density and the speed of sound for non-isotropic materials such as A value of 0.52 is adopted in the present work for the $K_{CFRP}$ coefficient [40].

*3.3 Vector flux model*

In this paragraph the methodology used to represent the debris environment surrounding the satellite so that it can be used in the computation of the survivability index is described. The methodology follows a novel approach in order to bypass the use of computationally expensive methods such as ray tracing, and at the same time have a sufficiently accurate description of the debris environment. The compromise is achieved characterising the debris fluxes with *vector flux elements*. Vector flux elements are entities used to reduce the debris flux data associated to a specific space sector around the satellite, generated by MASTER-2009, to a single vector element. To each vector element is then associated a value of the particle flux, of the velocity, and a direction. This in turn corresponds to the association of an entire set of particles to a specific value of the velocity, flux, and impact angle.

The procedure starts with the subdivision of the space around the satellite into a set of angular sector. The contribution to the particle flux inside each angular sector is summed together in order to generate the flux relative to the specific angular sector. The velocity and the impact direction are extracted from the velocity, impact elevation, and impact azimuth distributions. In order to have a clearer understanding of the procedure, we describe the procedure used to generate the vector flux elements with a simplified discretisation:

- Impact elevation: -5°, 0°, 5°.
- Impact azimuth: -90°, -75°, -60°, -45°, -30°, -15°, 0°, 15°, 30°, 45°, 60°, 75°, 90°.

Once the subdivision is decided, first the distribution of flux vs impact elevation is considered and subdivided in correspondence of the user specified subdivisions (Fig. 3). After the subdivision is performed, a value for the elevation angle for each of the elevation interval specified must be selected. Two options are possible: a *most probable value* or a *weighted average value*. The most probable elevation corresponds to the value of the impact elevation angle that has the highest flux in the interval specified. The weighted average elevation is instead given averaging the impact elevation values with weights given by the corresponding value of particle flux. When the subdivision is equal to the minimum possible width, 5 degrees in this case, the middle value is selected. At this point, the impact elevation associated to the vector flux element is determined.

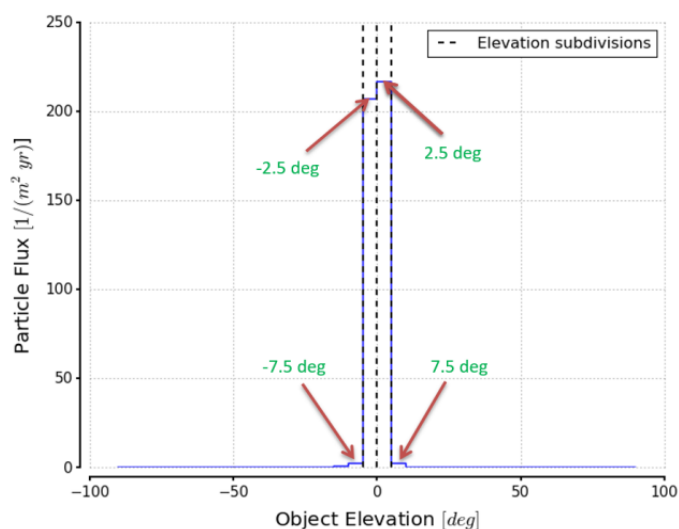

**Fig. 3: Differential flux vs impact elevation with distribution subdivisions (dashed line) and most probable values of the levation selected for each interval of the subdivision.**

The following step in the procedure allows the computation of the impact azimuth of the vector flux element. To do so, the 3D distribution of flux vs azimuth vs elevation is considered. First, the distribution is subdivided according to the elevation subdivisions. Each one of these subdivisions can contain multiple flux vs azimuth distributions generated by MASTER-2009. Such distributions are thus collapsed to produce one flux vs azimuth distribution for each elevation subdivision. These azimuth distributions are then subdivided following the user defined set of subdivisions (Fig. 4). The value of the azimuth associated to each interval is then computed. Following the same procedure used for the elevation distribution, there is the possibility to select between the *most probable* and *weighted average* value.

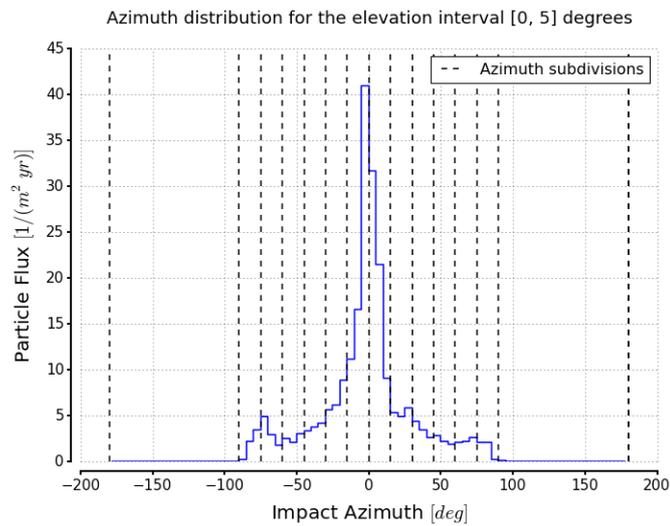

**Fig. 4: Subdivision example for the azimuth distribution corresponding to the elevation interval [0, 5] degrees.**

At this point, we have subdivided the debris environment surrounding the spacecraft in a set of angular sector and to each angular sector is associated a value of the impact elevation and of the impact azimuth that will serve to compute the direction of the vector flux linked to the correspondent angular sector. For the subdivision considered, we will have a total of 14 x 4 = 56 vector fluxes.

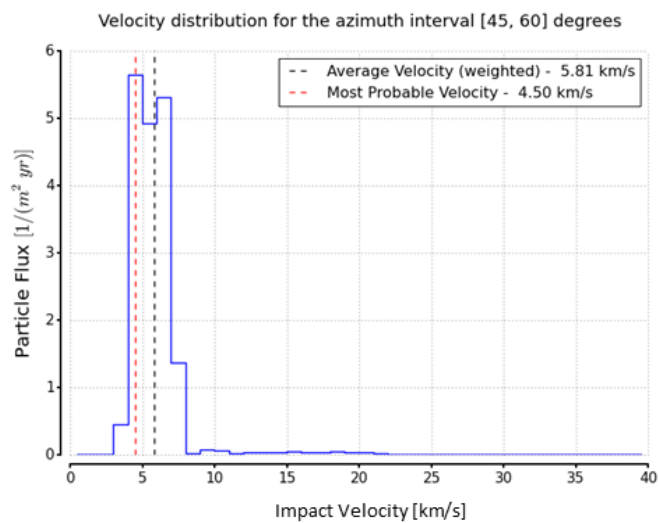

**Fig. 5: Velocity magnitude selection example.**

To complete the definition of the vector fluxes it is necessary to associate them a velocity magnitude. The process is analogous to the one followed for the impact azimuth and impact elevation but in this case the 3D distribution of flux vs azimuth vs velocity is considered. A set of flux vs velocity distributions is obtained. The number of distribution is again the same as the number of azimuth subdivisions (14 in this case). From each distribution, the magnitude of the velocity vector is extracted with possible choices again between the *most probable velocity* and the *weighted average velocity* Fig. 5).

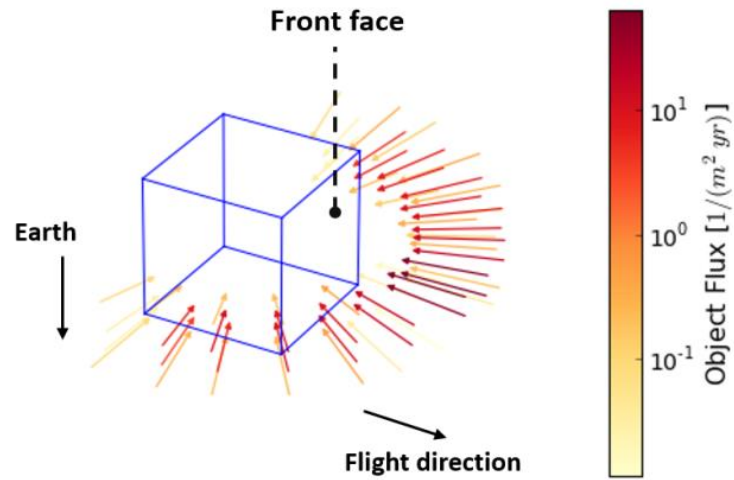

**Fig. 6: Vector flux representation for the example subdivision. The colour-map represents the flux magnitude of the vectors.**

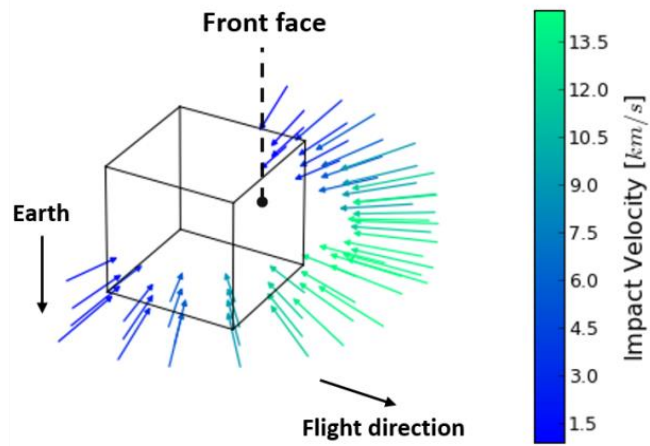

**Fig. 7: Vector flux representation for the example subdivision. The colour-map represents the velocity magnitude of the vectors.**

With this kind of procedure two type of simplifications have been introduced. First, it is considered that the velocity has a higher variability with respect to the impact azimuth rather than the impact elevation that is the reason why the distribution of flux vs azimuth vs velocity is used to compute the velocity magnitude of the vector fluxes. Second, using a unique value for the velocity, whether it is the most probable or the average, we are assuming that all the particles coming from a certain directions (angular sector) possess the same velocity. Finally, the set of vector fluxes with associated flux, velocity, and direction in the form of impact azimuth and elevation can be represented in a visual way as in Fig. 6 and Fig. 7.

*3.4 Spacecraft vulnerability*

The methodology used to assess the vulnerability of a satellite to debris impacts relies on a probabilistic approach where the penetration probability ($P_p$) on a certain surface is evaluated as a function of the particle flux, velocity, diameter, the projected area of the surface considered, and the mission duration in years. Once the vector flux elements are computed through the procedure described in Section 3.3, the areas of the spacecraft that are susceptible to particle impacts can be determined using a visibility role. Considering the object to be represented by a set $F$ of faces and a particle flux having a vector velocity equal to $v_i$, to verify if particles having that velocity vector can impact one of the panels in the set $F$ we check the velocity vector $v_i$ with respect to the panel normal $n_j$ with the following role:

$$\mathbf{n_j} \cdot \mathbf{v_i} < 0 \tag{35}$$

All the faces F representing the component are checked against all the vector flux elements following the same visibility role. At the current stage of the model no shadowing effects due to external components such as solar panels or antennas are considered and only the simplified role of Eq. (35) is used to check if a part of the spacecraft is hit by debris or not.

Then it is necessary to compute the probability of such impact. Assuming that debris impact events are probabilistic independent [45], it is possible to use a Poisson distribution (Eq. (36)) to compute the impact probability.

$$p(y) = \frac{\lambda^y}{y!} e^{-\lambda} \tag{36}$$

where $p(y)$ is the probability of $y$ impacts, and $\lambda$ is the expected number of impacts. From the Poisson distribution is possible to compute the impact probability, which is the complement of no particle impact ($y = 0$), considering that the expected number of impact can be expressed as:

$$\lambda_{j,i} = \varphi_i \cdot A_j^\perp \cdot t \tag{37}$$

where $\varphi_i$ (1/m²yr) is the *i-th* vector flux, $A_j^\perp$ (m²) is the projected area of the *j-th* face considered, and $t$ is the mission time in years. Thus, the impact probability of the vector flux $i$ onto the face $j$ is given by:

$$P_{j,i}^i = 1 - e^{-\varphi_i \cdot A_j^\perp \cdot t} \tag{38}$$

It is then necessary to compute the penetration probability on the panels. To do so, an alternative approach with respect to the standard one is adopted. The common procedure used to compute the penetration probability relies on the generation of many impact particles, whose diameter is then checked against the critical diameter computed using ballistic limit equations (BLEs) to assess if a penetration is occurred or not. As briefly introduced at the beginning of Section 3, the presented methodology bypasses the direct generation of the impacting particles. Ballistic limit equations are still adopted to compute the critical diameter using the velocity and direction associated with the vector flux elements together with the geometric and material characteristics of the panels. Once computed the critical diameter for the *i*-th vector flux onto the *j*-th the penetration probability is given by

$$P_{j,i}^p = 1 - e^{-\varphi_{c,i} \cdot A_j^\perp \cdot t} \tag{39}$$

where $\varphi_{C,i}$ is the particle flux with a diameter greater than the computed critical diameter. With the presented methodology, the computation of the critical flux for each vector flux element replaces the direct sampling of the debris particles. The critical flux can be extracted from the distribution of the cumulative flux vs diameter provided by MASTER-2009. The procedure is presented in Fig. 8, where the critical cumulative flux associated to the critical diameter is extracted from the distribution provided by MASTER-2009.

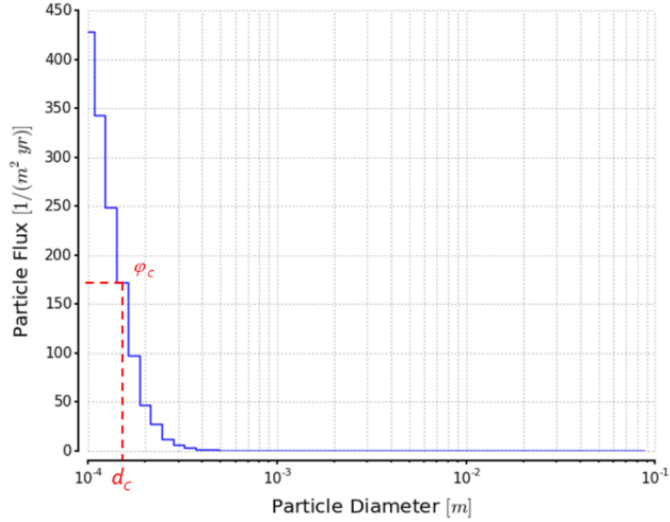

**Fig. 8: Critical flux computation methodology**

As the global distribution of cumulative flux vs diameter is used, the flux extracted is the overall flux for the entire range of azimuth and elevation, it thus cannot be directly used to compute the penetration probability relative to one of the vector fluxes. To each vector flux, in fact, is associated a value of the particle flux that is dependent upon the directionality, i.e., impact elevation and impact azimuth, which is a fraction of the total flux. It is here assumed that the distribution of the particles diameter is uniform with respect to the impact direction. With this assumption, the critical flux associated to a vector flux element is considered as a fraction of the overall critical flux. If $\varphi_{TOT}$ is the total debris flux (for example the value between 400 and 450 in Fig. 8) and $\varphi_C$ is the overall critical flux computed, the critical flux relative to the considered vector flux element can be expressed as:

$$\varphi_{C,i} = \varphi_i \cdot \frac{\varphi_C}{\varphi_{TOT}} \tag{40}$$

Finally, the penetration probability on a structure composed by multiple panels can then be computed as follows:

$$P^p = 1 - \prod_{j=1}^{N_{panels}} \left( \prod_{i=1}^{N_{fluxes}} \left(1 - P_{j,i}^p\right) \right) \tag{41}$$

with $N_{fluxes}$ total number of vector fluxes elements and $N_{panels}$ total number of panels composing the structure.

*3.5 Computational model*

A brief overview of the inputs necessary to the model and the possible options available to the user together with the output obtained are here provided.

*3.5.1 Inputs*

The first necessary inputs are the flux distributions obtained from MASTER-2009. As it has been described, four distributions are used and the user can decide freely the orbit characteristics and the distribution ranges and bins to be provided to MASTER before running it. After MASTER produces the text files with the discretised distributions, they can be loaded into the code in order to start the analysis. Then the subdivisions used to generate the vector fluxes have to be provided as it has been illustrated in section 3.3. The choice of where to subdivide the impact azimuth and elevation intervals is arbitrary but the lowest subdivision width does not have to be lower than the minimum subdivision specified in MASTER-2009. Another fundamental input to the model is the geometrical description of the structure to be analysed. The geometry has to be provided in the form of a list containing for each panel the outward normal, the surface area, and the thickness of the panel. Along with the geometrical information also the material characteristics and the failure modality (perforation, detached spall, or incipient spall) has to be provided. The model thus gives the possibility to the user to define different materials, geometries, and failure modalities for different part of the structure. Finally, the mission duration in years and the density of the debris has to be provided. The latter option allows the user to distinguish between man-made debris and micrometeoroids as the first ones have higher density (2.8 g/cm$^3$) [43] with respect to the second ones (1.0 g/cm$^3$) [43].

### 3.5.2 Outputs

The output of the program is the penetration probability on each panel constituting the structure and the overall penetration probability. An example of the output for a 800 km altitude and 98° inclination orbit, for a one-year mission on a 1 m side length cubic structure of aluminium with thickness equal to 6 mm is given in Table 6

**Table 6: Penetration probability on each face of a 1m side, 6mm thick cube, and total penetration probability for the Orbit 1 case.**

| Face | Penetration Probability |
|---|---|
| Front | 0.00808173704 |
| Back | 0.00000002708 |
| Left | 0.00117693069 |
| Right | 0.00109870168 |
| Top | 0.00020032434 |
| Bottom | 0.00019310123 |
| **Total Penetration Probability** | 0.01072704267 |

### 3.6 Model validation

The presented model has been compared with the MIDAS module of the ESA software suite DRAMA [46]. MIDAS is able to perform a vulnerability analysis on oriented surfaces using debris fluxes extracted from ESA MASTER-2009. The user has to specify the star and ending epoch of the simulation, the orbital elements of the operational orbit, and the thickness and density of the material of the panels. To compare the two codes a cubic shaped structure has been chosen with characteristics summarised in Table 7. In order to accurately compare the two codes, the same BLE has been used, corresponding to the Cour-Palais single wall damage equation as specified in Gelhaus, Kebschull, Braun, Sanchez-Ortiz, Parilla Endrino, Morgado Correia de Oliveira and Dominguez-Gonzalez [46].

**Table 7: Characteristics of the test structure.**

| Shape | Cube |
|---|---|
| Dimensions | 1 X 1 X 1 m |
| Thickness | 2 mm |
| Material | Aluminium 6061-T6 |
| Damage equation | Cour-Palais single wall |

Two mission scenarios have been selected for the comparison. Their characteristics are summarised in Table 8. The first scenario is a one-year mission in a typical sun-synchronous orbit. The second scenario is a one-year mission in the International Space Station (ISS) orbit. The scenarios has been selected has representative to test the reliability of an impact vulnerability code. The first orbit is characterised by very high debris fluxes, with high impact velocities, especially focused on the leading face of the structure. The second orbit, instead, is characterised by lower fluxes, with lower velocity impacts. The impacts are also more evenly distributed between the lead, left, and right faces of the structure.

Table 8: Mission scenarios characteristics.

| Orbit Type | h (km) | i (deg) | e | Start epoch | End epoch |
|---|---|---|---|---|---|
| SSO | 802 | 98.6 | 0.001 | 01/01/2016 | 01/01/2017 |
| ISS | 400 | 51.64 | 0.001 | 01/01/2016 | 01/01/2017 |

Table 9 shows the results for the sun-synchronous orbit test case. The number of impact and number of penetrations for each face of the structure are compared for the two codes. The totals are also summarised at the bottom of the table. The results show a very good agreement for both the total number of impact and the total number of penetrations, with only a very small difference between the two models. The face-by-face comparison also shows a very good agreement. Two differences that are more noticeable can be spotted in the number of impacts for the Space and Earth faces and in the number of penetrations for the Trail face. The first difference can be explained with the approach followed by the described survivability model, where a discretised representation of the debris fluxes is used, with weighted average values for the direction and velocity of the vector flux elements. For the second difference, it can be explained remembering the procedure used to compute the critical flux (Section 3.4). In the case of the trailing face, the velocity and fluxes are very low, thus generating large values for the critical diameters. If the critical diameter is greater than the upper limit for the distribution extracted from MASTER-2009 (0.01 m), the value of the flux corresponding to this upper limit is taken. However, the order of magnitude of the five times lower than the number of penetrations of the leading face, and thus has a negligible influence on the final result.

Table 9: Comparison with DRAMA for a sample Sun-Synchronous orbit (SSO).

| Test SSO | Number of impacts | | Number of penetrations | |
|---|---|---|---|---|
| Face orientation | MIDAS | Survivability Model | MIDAS | Survivability model |
| Lead | 69.473 | 69.47311217 | 2.88E-01 | 2.77E-01 |
| Space | 0.48114 | 1.078412829 | 1.71E-05 | 2.12E-05 |
| Trail | 0.032326 | 0.032675576 | 7.60E-11 | 1.03E-06 |
| Earth | 0.54294 | 1.259677243 | 1.73E-05 | 1.99E-05 |
| Right | 19.196 | 19.17890039 | 1.03E-02 | 6.75E-03 |
| Left | 21.953 | 21.97517311 | 1.00E-02 | 7.64E-03 |
| **Total** | 111.678406 | 112.9979513 | 0.309034425 | 0.291390421 |

Table 10 summarises the same comparison for an ISS-like orbit. Even in this case the two models compare well, with the results closely following each other. In this case, the difference in the number of impacts for the Space and Earth faces are smaller, and all the other faces compares very well, up to two decimals. For the number of penetrations, the results are still comparable but the agreement is reduced. Again, this is probably due to the way the vector flux elements are generated. In fact, the fluxes characteristics of ISS-like orbits have a less directional behaviour than the one in SSO orbits that is the fluxes and the velocities are more evenly distributed over the impact directions. The sampling methodology used to generate the vector flux elements is less capable of capturing this behaviour.

Table 10: Comparison with DRAMA for the ISS orbit.

| Test ISS | Number of impacts | | Number of penetrations | |
|---|---|---|---|---|
| Face orientation | MIDAS | Survivability Model | MIDAS | Survivability model |
| Lead | 1.5241 | 1.528180145 | 1.21E-02 | 8.03E-03 |
| Space | 0.088823 | 0.115116158 | 2.62E-06 | 1.20E-05 |
| Trail | 0.04817 | 0.0479896 | 5.69E-06 | 5.74E-06 |
| Earth | 0.092399 | 0.116110808 | 9.85E-06 | 1.21E-05 |
| Right | 1.638 | 1.635195747 | 3.69E-03 | 4.85E-03 |
| Left | 1.125 | 1.127619142 | 2.74E-03 | 3.01E-03 |
| **Total** | 4.516492 | 4.570211601 | 0.01518409 | 0.015922959 |

*3.7 Survivability criterion*

To evaluate and compare each solution using the developed survivability model, a survivability criterion has been introduced. The criterion used is defined as the probability of no-penetration (*PNP*), which can be expressed as:

$$PNP = 1 - P_p \tag{42}$$

where Pp is the penetration probability as computed with Eq. (41). A value of 0 of the *PNP* index corresponds to a completely vulnerable structure, whose probability to be penetrated by a debris during the mission lifetime considered and given its characteristics is 100%. A value of 1 instead correspond to a fully protected configuration.

**4   Demisability and survivability analysis of design-for-demise solutions**

Design-for-demise options are a tool to design a spacecraft compliant with the casualty risk regulations. Among the specific methods employed in designing spacecraft parts to demise, the following can be identified [7, 9]:
- Use of a different material: re-designing a spacecraft using a more demisable material;
- Use of multiple materials: replacing a single non-demisable material with different demisable materials while still maintaining structural integrity;
- Changing the shape: changing the shape of an object can change its area-to-mass ratio and enhance the demisability;
- Changing the size: change the dimensions of a component to modify its area-to-mass ratio;
- Changing the thickness: results in a change of the mass of the object thus altering the area-to-mass ratio;
- Changing the component location: locating a component close to the exterior structure of the spacecraft can expose the object to the ablative environment earlier than a component located in the inner part of the spacecraft;
- Promoting components early break-up: the early break-up could produce a more prolonged exposure to the re-entry heat load thus promoting the demise of a component.

These design solutions, affects the spacecraft and its components and, in general, they aim at producing structures with an increased area-to-mass ratio to enhance the demisability. As a result, lighter, thinner, and more exposed structures are designed. These resulting design-for-demise structures, given their new characteristics, may be more vulnerable to the impact from space debris. In the following paragraphs, the introduced design-for-demise solutions, except for the component location, are analysed using the developed demisability and survivability models (Section 2 and Section 3). The current state of development of the model, in fact, does not allow such an analysis to be carried out. However, it is intention of the author to dedicate future efforts to the subject.

It is clear that the problem in exam is extremely complex. Multiple parameters influence both the demisability and the survivability of a spacecraft and its components. Some of these parameters are related to the system design of the spacecraft. The design-for-demise solutions belong to this category as they directly affects the manufacturing of specific parts of the spacecraft. In addition, mission related parameters such as the operational orbit of the satellite and its disposal strategy can influence the resulting survivability and demisability. Moreover, the dependency of the demisability and the survivability upon these parameters is usually non-linear, thus making difficult to generalise the results obtained. In fact, the selection of specific design-for-demise options and their consequences on the mission design and requirements is very much a mission dependent consideration.

However, it is still possible to study such design options. Understanding in which way they influence the demisability and the survivability. Evaluating if a specific solution is more effective than others are, or if the effects of a solution can independently affect the demisability or the survivability. In order to study such behaviour and evaluate the design-for-demise options, the single options are analysed independently, keeping constant the other parameters. In this work, the analysis each of the solutions consists in the evaluation of both the demisability and the survivability over a range of initial conditions so that the variability of the indices can also be considered. In this way for each design solution it is possible to evaluate average effect, measured using the developed indices. Alongside the average effect, also the standard deviation can be computed to provide an evaluation of how variable is the effect of the specific design solution as a function of the initial conditions. Before examining the design-for-demise options by taking into account their average behaviour over a range if initial conditions, it is important to understand which of these conditions is actually more influential to the problem in exam. A sensitivity analysis as thus been carried out over the initial condition provided to the demisability and survivability models.

*4.1 Sensitivity analysis methodology*

A sensitivity analysis has been performed to understand the inputs that mostly affect the demisability and survivability criteria presented and which parameters can instead be neglected in future studies. The procedure selected to perform the sensitivity analysis relies on the Sobol method [47], which can be classified among the variance-based sensitivity methods. Variance-based methods [48, 49] exploit the decomposition of the variance of the model output into terms depending on the input factors and their interactions. They allow the computation of the first order and total order effects of the input parameters, as well as the mutual interactions between them. Such methods are extremely versatile [47, 49] and can be used with many different models. Using a Monte Carlo based sampling procedure they are suitable to be used with non-linear models and models whose input parameters are not directly correlated [47, 50]. In the paper we use the Sobol methodology coupled with the Saltelli sampling technique [51, 52]. In the Saltelli sampling a number $N = (2k + 2) \cdot n$ simulations has to be ran, where $k$ is the number of input parameters, and $n$ is the sampling size. Throughout the paper, the number of sample $n$ used is equal to 2000.

*4.2 Representative component for the sensitivity analysis*

To perform the described analysis it was decided to consider the effects of the design-for-demise options on a reference internal component. The choice of the component has to consider both the characteristics of the demisability and the survivability analysis, taking into account a component whose analysis is interesting for both aspects. Following this consideration, we decided to select a tank as reference component. Tanks are in fact sensitive components for the demisability as they usually survive re-entry [53-55]. At the same time, it is important to ensure adequate protection and resistance of tanks from debris impact as, being pressurised, they are particularly susceptible to ruptures and leakage [39]. The characteristics of the selected representative tank are summarised in Table 11.

**Table 11: Characteristics of the reference tank.**

| Parameters | Ranges |
|---|---|
| Mass | 15 kg |
| Shape | Cylinder |
| Length | 0.896 m |
| Diameter | 0.6 m |

| Material | Steel AISI-316[†] |
|---|---|

The characteristics of the tanks have been derived from the data available for the MetOp-A mission [56-58]. The mission was selected to be representative of medium to large LEO satellites, which are the most interesting to analyse for design-for-demise options. Moreover, the satellite belongs to the sun-synchronous region that is the most exploited LEO region, making it a good candidate to be representative of many mission scenarios. Knowing the amount of propellant stored by MetOp-A (316 kg), and the tank manufacturer [57], a suitable tank was selected, considering a filling factor of 0.4 and a number of tanks equal to four.

The presented analyses have some limitations that are mainly a consequence of the simplified nature of the models used. At the current stage of development, it is not possible to model internal components. This allow considering only single simplified components, preventing the analysis from modelling realistic spacecraft configurations. As such, the break-up of internal components cannot be modelled, as well as the shielding effect of the external structures on the internal components. This last point definitely affects the absolute value of the survivability index. In fact, with the presence of the external structure, the penetration probability on internal components is greatly reduced in magnitude. However, it was considered that the results obtained with an object directly exposed to the debris fluxes could be extended also to internal components. As mentioned, this is not valid in terms of the absolute values of the PNP index but is still valid in relative terms. It is in fact reasonable to assume, for example, that changes to a component design that increase its survivability, evaluated when it is directly exposed to the debris fluxes, will still produce an increase of the survivability when it is protected also by the external structure.

### 4.3  Sensitivity to mission characteristics

When considering the demisability and the survivability models as described in Section 2 and 3, it is necessary to provide them some parameters in order to perform the simulations. For the demisability, the initial conditions at re-entry have to be provided in the form of initial altitude, relative velocity, flight path angle, longitude, latitude, and heading angle. For the survivability, on the other hand, the operational orbit has to be provided, which consists of the altitude, and inclination of the orbit. In addition, the mission lifetime in years has to be specified. These are the main characteristics that influence the re-entry of a spacecraft and its vulnerability to debris impact. As such, these are the parameters taken into account in the sensitivity analysis presented. The reference component selected for the analysis is the tank described in Table 11.

#### 4.3.1  Demisability sensitivity to re-entry inputs

Here, the sensitivity of the demisability is shown varying all the parameters defining the initial conditions at the re-entry interface. The values ranges adopted in the sensitivity analysis are summarised in Table 12. All the variables in Table 12 are varied uniformly inside the specified ranges. Such choice reflects the fact that the demise of components is only studied after the break-up of the main spacecraft body happens. The standard value for the break-up altitude is 78 km [46, 59]; however, this value can vary according to the satellite characteristics and entry scenario. Despite the difficulties in predicting the actual break-up altitude, it is possible to consider its variation in a sensitivity analysis and study its contribution to the demisability. The selected altitude range reflects a wide range of possible re-entry scenarios. The longitude, latitude, and heading angle ranges have been selected to take into account all the possible re-entry locations and orientations. The range of velocity is considerably wide, in order to take into account for different re-entry scenarios and characteristics of the spacecraft. In fact, the velocity at break-up is influenced by the slope of the entry trajectory, the initial velocity, and the aerodynamics of the satellite main body. The range for the flight path angle was selected as representative of uncontrolled re-entry, either from a slowly decaying orbit or from a disposal orbit directly targeting a re-entry.

---

[†] The original material of the tank is the titanium alloy Ti-6Al-4V. The decision to change the tank material was made in order to have a greater variety of results for the different initial conditions. Titanium alloys, in fact, rarely demise and/or reach the melting temperature.

**Table 12: Ranges of parameters for the sensitivity analysis of re-entering components.**

| Parameters | Ranges | |
|---|---|---|
| Longitude | 0° | 360° |
| Latitude | -90° | 90° |
| Altitude | 60 km | 100 km |
| Entry velocity | 7.1 km/s | 8.1 km/s |
| Flight path angle | -5° | 0° |
| Heading angle | -90° | 90° |

The results of the sensitivity analysis are shown in Fig. 9. The most influential parameters are the altitude and the velocity at the break-up. All the other parameters have a considerably lower Sobol index (the altitude contribution is more than ten times the one of the flight path angle). In particular, latitude and longitude have an almost negligible sensitivity with respect to the demisability. It is nonetheless important to mention that latitude and longitude become important in case the casualty risk on ground is considered. In this case, in fact, they take into account the variability of the population on Earth, which is one of the main parameters affecting the computation of the casualty risk. Both the first order effects and the total order effects are presented in the analysis. The first order effects measure the effect that changing a single variable has on the output. In a Sobol sensitivity analysis, this is the contribution to the output variance caused varying a single parameter, but averaged over the variations in other input parameters. The total order effects instead measure the contribution to the output variance of a variable, including all variance caused by its interactions, of any order, with any other input variables. It is interesting to observe that the first and total order effects are of comparable magnitude. This means that in the re-entry of a spacecraft component, the value of a parameter directly influences the demisability even with different combinations of the other parameters.

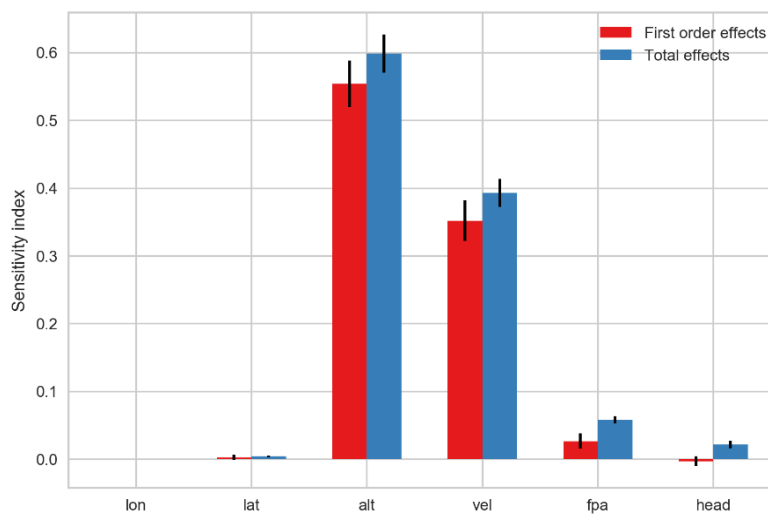

**Fig. 9: First order and total order effects for the re-entry sensitivity analysis of components.**

*4.3.2 Survivability sensitivity to operational conditions*

To study the sensitivity of the survivability for a component, the altitude, the inclination, and the mission lifetime have been considered. The values ranges adopted in the sensitivity analysis are summarised in Table 13. The study focuses on the LEO region; as such, the maximum altitude was set to 2000 km. The maximum mission operational lifetime has been set to 15 years as it was considered a reasonable upper value for the duration of a mission. The specified time range only includes values for the operational life of the satellite, without considering the disposal time needed by the satellite to re-enter into the atmosphere after decommissioning. In this latter case, a larger time span would be needed to include the entire lifetime of the mission. The inclination range instead takes into account all possible orbit options.

Table 13: Ranges of parameters for the survivability sensitivity analysis.

| Parameters | Ranges | |
|---|---|---|
| Altitude | 700 km | 2000 km |
| Inclination | 0° | 180° |
| Mission lifetime | 1 yr | 15 yr |

The results of the sensitivity analysis are presented in Fig. 10. As expected, the altitude has the highest impact on the survivability analysis. This is due to the variation of the debris fluxes as a function of the altitude, where the highest concentrations are in correspondence of specific altitude ranges such as the bands between 600 km and 900 km and between 1400 km and 1500 km [1, 2, 60].

A slightly less expected result is the influence of the orbit inclination, being almost half the sensitivity index of the altitude. Orbit inclination can affect the penetration probability on a structure because different flux concentrations can be present ad different inclinations and because of the flight direction of the spacecraft with respect to space debris. Retrograde orbit are in general more dangerous than direct orbits as they generates impacts with a larger relative velocity. This is probably because orbit inclination becomes a more important factor only for those orbits with higher debris fluxes (e.g. sun-synchronous orbits). As such, over the wide range of altitude considered, the debris density variation with the altitude becomes a more determinant factor than the inclination. Nonetheless, the contribution of the orbit inclination is still important and is thus considered for the remaining of the analysis. Finally, the mission lifetime has clearly a strong influence on the survivability. As we could expect, the longer the mission the higher is the probability of a component being damaged by debris impacts (Fig. 10).

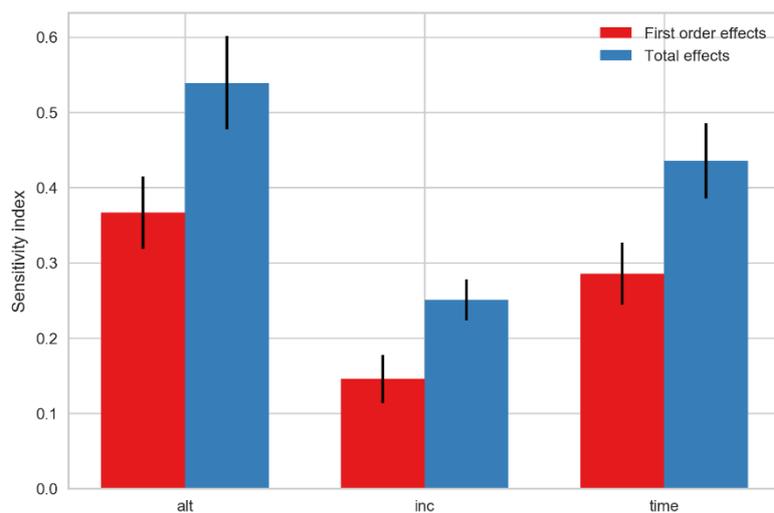

**Fig. 10: First order and total order effects for the survivability sensitivity analysis of components.**

*4.4    Effects of the design-for-demise solutions*

Once the main parameters affecting the demisability and the survivability have been identified, it is possible to focus the attention towards the effect that the design-for-demise solutions have, on average, on the demisability and the survivability. To do so, the identified design-for-demise options are considered singularly, varying them while keeping constant the other parameters. In this way, their effect is isolated from the others and the variation of the parameter can be taken into account.

Following the results obtained in Section 4.3, the effect of the design-for-demise options is studied varying the three identified most influential initial conditions (i.e. altitude, entry velocity, and flight path angle) for each of the solutions available in the design-for-demise options. For each option, a Monte Carlo simulation was performed varying the initial conditions inside the altitude, velocity, and flight path angle ranges (Table 12) for the demisability, and the altitude, inclination, and mission time ranges for the survivability (Table 13). Again, the Saltelli sampling methodology was used with 2000 samples. In this way, it is possible to perform a more general analysis of the impact of the design-for-demise solutions as they are evaluated over the entire range of possible initial conditions.

The obtained plots represent for each solution the average value of the demisability and of the survivability indices over the set of Monte Carlo simulations, together with the standard deviation of set of solutions. In addition, to the plot is added the percentage of solutions above a threshold value of demisability and survivability for each solution. For the demisability, the selected threshold is a value of the *LMF* index equal to 0.9. For the survivability, the equivalent threshold is a value of the PNP index equal to 0.99. This percentage is directly related to the quality of the solutions. A higher percentage corresponds to a solution more likely to generate a more robust design in terms of demisability and survivability over a wide range of initial conditions. In a preliminary design phase, this is an important aspect as many aspect of the mission design can still be modified.

*4.4.1 Changing the component material*

The first design-for-demise option considered it the change of the material of the component. The materials taken into account are the aluminium alloys 6061-T6 and 7075-T6, the stainless steels AISI-304 and AISI-316, the Inconel-601 alloy, the titanium alloy 6Al-4V, and the graphite epoxy. Titanium and stainless steel alloys are common materials used in the manufacturing of spacecraft tanks. Aluminium alloys have been also considered as they can be used to manufacture tanks and they are being studied by ESA Clean Space as demisable substitutes of currently non-demisable titanium tanks [10, 11]. Graphite epoxy has been selected as representative of composites solutions. Two options for the graphite epoxy are taken into account, as there is a significant discrepancy between the models used in different software to take into account such material. According to the model used in DAS, the graphite epoxy is extremely demisable as it is considered to char as it reaches its melting temperature. However, according to other studies, graphite epoxy and composites in general result very resistant to the ablation process, usually surviving the re-entry process [61, 62]. This is especially due to the properties of carbon fibres, which are very temperature resistant

Two different cases were considered for here. Components with a constant thickness (2 mm in this analysis) and components where the thickness is varied to coincide with the minimum allowable thickness sustainable by the tank. In the first case, together with the thickness, the dimensions are kept constant and correspond to the one specified in Table 11. In the second case, the dimensions are constant, and the thickness is varied in order to maintain the structural integrity. The minimum allowed thickness is computed using Eq. (43) [63].

$$t_{s,\min} = \frac{K_{sf} \cdot p_{\max} \cdot r}{\sigma_u} \tag{43}$$

where $\sigma_u$ is the ultimate strength of the material, $r$ is the radius of the cylinder, $p_{max}$ is the maximum operating pressure (assumed equal to 4 MPa) and $K_{sf}$ is a safety factor (assumed equal to 1.5). This analysis is meant to compare the performance of feasible tanks configuration for the different material considered. It is important to underline the fact that changing the material of tanks can also have a considerable influence on the mass budget of the mission. Table 14 summarises the weights of the tanks for each material option used in the analysis and for both the cases considered.

Table 14: Summary of the mass impact of tank solutions as a function of the material and case considered.

| Tank material | Tank mass (kg) Constant thickness | Tank mass (kg) Minimum thickness |
|---|---|---|
| Al-6061T6 | 12.161 | 26.448 |
| Al-7075T6 | 12.493 | 27.169 |
| AISI-304 | 35.412 | 39.985 |
| AISI-316 | 35.981 | 40.627 |
| Inconel-601 | 36.117 | 40.781 |
| Ti-6Al4V | 19.889 | 14.244 |
| Graphite-epoxy-1 | 7.038 | 9.551 |
| Graphite-epoxy-2 | 6.95 | 9.432 |

Fig. 11 and Fig. 12 show the results for both cases and for different materials. In both graphs, the top plot represents the behaviour of the Liquid Mass Fraction, and the bottom graphs represent the probability of no-penetration. The black points and the bars represent the average values and standard deviations of each Monte Carlo simulations. The blue lines shows the percentages of solutions for each material case whose *LMF* index is higher than 0.9. The green lines represent the percentages of solutions for each material case whose *PNP* index is higher than 0.99.

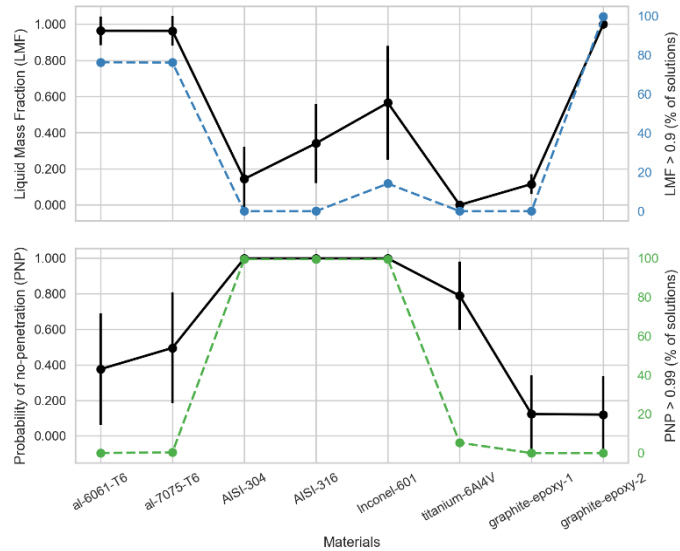

**Fig. 11: Variation of the average value and standard deviation of the *LMF* and *PNP* indices for different materials and a constant thickness configuration (black lines). Percentage of solutions over a 0.9 *LMF* index (blue line) and over a 0.99 *PNP* index (green lines).**

From the presented analysis, it is first possible to observe the considerable difference in both the demisability and survivability index that is obtained when just a change in the material is introduced. Taking a closer look to the demisability, we can give a ranking to the materials based on their *LMF* index results. Aluminium alloys and the graphite-epoxy-2 have the highest demisability. The results associated to the graphite-epoxy-2 are here presented for completeness; however, they constitute a very optimistic representation of the behaviour of composites material, which has been proved not to be very accurate [62]. The main destructive re-entry codes, such as SESAM, SCARAB, and SAM use a much more conservative approach that can be best represented by the graphite-epoxy-1 case.

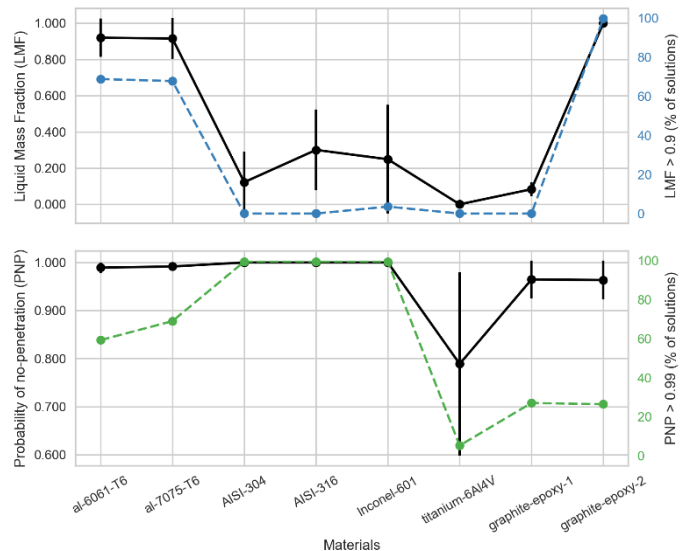

**Fig. 12: Variation of the average value and standard deviation of the *LMF* and *PNP* indices for different materials and a minimum allowable thickness configuration (black lines). Percentage of solutions over a 0.9 *LMF* index (blue line) and over a 0.99 *PNP* index (green lines).**

Aluminium alloys, as expected, prove to be very effective in improving the demisability of the analysed components, also resulting in robust design, given the high percentage of solution with a *LMF* index over 0.9. On the other end, materials such as titanium and graphite-epoxy-1 are extremely difficult to demise, producing no solutions with a demisability above 0.9. In the middle between these two extreme behaviours, there is the group represented by the stainless steels and the Inconel alloy. Such solutions can reach a LMF index around half the one of aluminium. However, only for the Inconel alloy solutions, the demisability can reach the level of 0.9. As such, among this group of material the Inconel alloy proves to be the most demisable one. Titanium, as expected, generated no demisable solutions in both cases.

Looking at the other side of the problem that is the correspondent change in survivability introduced when changing the material, it is evident that the particular case in exam makes a considerable difference. In Fig. 11, where the thickness is kept constant to 2 mm (a typical value for spacecraft component [64] the *PNP* changes significantly between the different materials. The clear favourite in terms of resistance to debris impact are the stainless steels and the Inconel alloys. Follow the aluminium alloys with a *PNP* index half the one of the stainless steels, to terminate with the graphite epoxy. The behaviour described by Fig. 11 can be considered as the absolute scale of value for the survivability of the considered materials as the main characteristics influencing the PNP index (dimensions and thickness) are kept constant.

In Fig. 12, the results are more directly applicable to tanks, where the thickness has to have a value greater than a minimum allowable value that will ensure structural integrity. In this case, it is possible to observe, that the aluminium alloys and the graphite epoxy have a *PNP* closer to the one reached by the stainless steel solutions. Still the difference is still in one order of magnitude as the *PNP* for the aluminium alloy solutions is in the order 0.99, whereas the *PNP* for stainless steel solutions is in the order of 0.999. A behaviour that is more clearly deduced looking at the probability of having solutions with *PNP* greater than 0.99 (green dashed line). Titanium solutions in this case are much more vulnerable to debris impacts because, given the high ultimate strength of titanium alloys (950 MPa), the solutions considered have the lowest thickness (about 1.5 mm), thus making it about 20% more vulnerable. Aluminium solutions, on the other hand, have a thickness of about 4.4 mm whereas the stainless steel solutions of about 2.1 mm.

From the results in Fig. 12, it possible to deduce that by only changing the material of commonly used tank configuration to an aluminium alloy, it is very likely to achieve a demisable solutions or at least a solution that will only require other minor changes to be completely demisable. Under the constraint of maintaining structural integrity, such solution also proves to have a very good survivability, almost comparable to the one of a stainless steel solution of comparable resistance.

However, given the thickness is more than doubled such solution would also be heavier (Table 14). In any case, as it is possible to observe, the necessity to consider more alternatives and carry out trade of solution, also considering the other requirement and constraints of a mission is of primary importance.

*4.4.2 Changing the component thickness*

Another option to act on the demisability of a component is to change its thickness. This solution can be used in combination with other design-for-demise options but can also be useful in all those cases in which other options such as changing the material or the dimensions of a component are not viable. To present the dependence of the demisability and of the survivability upon the thickness variation, all the characteristics of the reference tank have been kept constant, except for the thickness itself and the material. Different materials are taken into account, and the results for the aluminium alloy 7075-T6 and the stainless steel AISI316 are presented in Fig. 13 and Fig. 14 respectively. The thickness has been varied from 0.5 mm to 10 mm in steps of 0.5 mm.

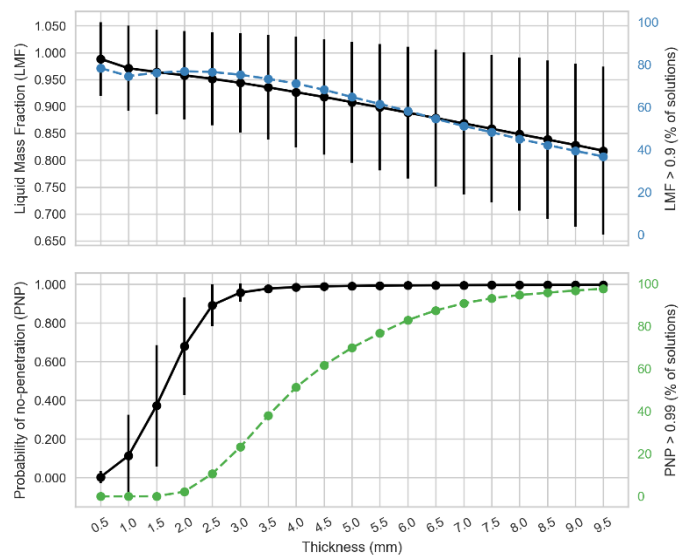

**Fig. 13: Variation of the average value and standard deviation of the *LMF* and *PNP* indices for variable thicknesses of aluminium alloy componenents (black lines). Percentage of solutions over a 0.9 *LMF* index (blue line) and over a 0.99 *PNP* index (green line).**

It is possible to observe how the behaviour of the demisability and survivability indices is different with a variation of the thickness. The liquid mass fraction, in fact, varies smoothly, constantly decreasing for the aluminium (Fig. 13), and reaching a maximum around 2 mm for the stainless steel (Fig. 14). On the other end, the probability of no-penetration has a steeper trend, with a considerable variation in a small range of thicknesses and a flattening afterwards. As expected, an increase in the thickness translates in a lower demisability for the aluminium alloys. As the thickness increases also the standard deviation of the solutions increases indicating a higher dependence of the outcome on the initial conditions for less demisable solutions. In a similar but opposite fashion, the standard deviation of the solutions for the survivability increases with decreasing thickness. In fact, as the vulnerability of the solution increases, the orbit selected becomes more influential.

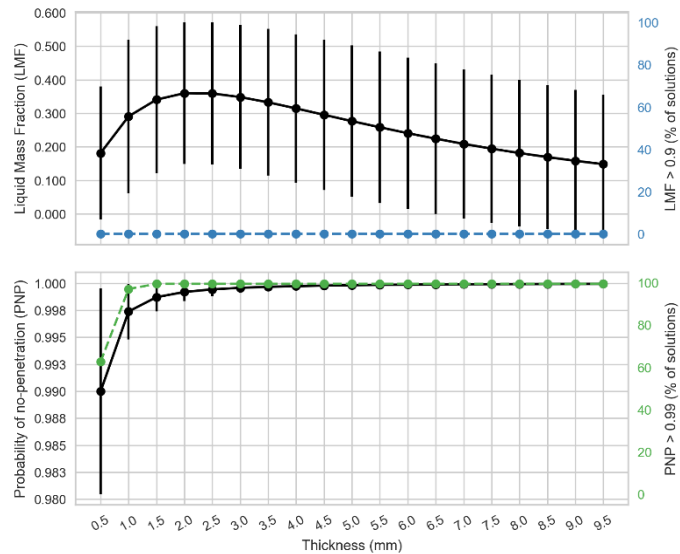

**Fig. 14: Variation of the average value and standard deviation of the *LMF* and *PNP* indices for variable thicknesses of stainless steel componenents (black lines). Percentage of solutions over a 0.9 *LMF* index (blue line) and over a 0.99 *PNP* index (green line).**

In general, it is possible to observe that aluminium alloy solutions are more robust on the demisability side, granting a high *LMF* index over a wide range of thicknesses. The stainless steel solutions, on the other hand, are more robust on the survivability side, generating very resistant structure even with small thicknesses. In fact, the percentage of solutions with a *PNP* index greater than 0.99 is over 90% in all the cases except for the smaller thickness (0.5 mm).

From the presented plots is possible to deduce that the thickness has a more direct influence on the survivability rather than on the demisability. Changing the thickness results in orders of magnitude changes in the probability of no-penetration. This is strikingly evident for the aluminium alloys but is also true for the stainless steel case. Moreover, the requirements over the probability of no-penetration are usually very strict, aiming at components with a *PNP* index below 1%.

The influence on the demisability is instead more gradual. For the stainless steel case (Fig. 14), despite a 20% change in the *LMF* index can be obtained varying the thickness, in no case such difference produces solutions with a demisability over 0.9. Consequently, a change in the thickness would need to be coupled with other design-for-demise solutions to achieve a substantial effect. In the aluminium alloy case, instead the *LMF* index remains always high and with a good percentage of solutions over a 0.9 *LMF* index. Therefore, a high demisability can be maintained coupling the thickness change with small changes in the other design-for-demise options.

*4.4.3   Changing the component dimensions*

Changing the component dimensions is an option that can be implemented in order to change the area-to-mass ratio of the component, trying to increase its demisability. However, when the dimensions change, also the survivability of the component changes. For example, larger components have a greater exposed area to debris flux, which results in a higher vulnerability of the component itself. Changing the dimension of a component can also result in the subdivision of a large component into smaller parts. For example, a large battery assembly can be subdivided into more than one box, making it more demisable. The same procedure can be carried out with tanks, dividing the amount of propellant into more vessels. This last case is the one considered here to show the dependence of the demisability and the survivability with respect to changing the dimensions of a component.

Starting from the tank configuration of Table 11 the range of dimensions has been varied as if the number of tanks ranges from one to ten vessels. The aspect ratio of the tank is kept constant. The thickness is instead varied according to Eq. (43) in order to compare realistic configurations. The corresponding dimensions of the tanks for the different configurations is summarised in Table 15 and Table 16.

**Table 15: Radius and thickness values for the configurations with different number of aluminium alloy tanks.**

| Number of tanks | 1 | 2 | 3 | 4 | 5 | 6 | 7 | 8 | 9 | 10 |
|---|---|---|---|---|---|---|---|---|---|---|
| Radius (m) | 0.366 | 0.290 | 0.254 | 0.230 | 0.214 | 0.201 | 0.191 | 0.183 | 0.176 | 0.170 |
| Thickness (mm) | 6.946 | 5.513 | 4.816 | 4.376 | 4.062 | 3.823 | 3.631 | 3.473 | 3.339 | 3.224 |

**Table 16: Radius and thickness values for the configurations with different number of stainless steel tanks.**

| Number of tanks | 1 | 2 | 3 | 4 | 5 | 6 | 7 | 8 | 9 | 10 |
|---|---|---|---|---|---|---|---|---|---|---|
| Radius (m) | 0.366 | 0.290 | 0.254 | 0.230 | 0.214 | 0.201 | 0.191 | 0.183 | 0.176 | 0.170 |
| Thickness (mm) | 3.589 | 2.848 | 2.488 | 2.261 | 2.099 | 1.975 | 1.876 | 1.794 | 1.725 | 1.666 |

The results are presented in Fig. 15 and Fig. 16 for aluminium alloy and stainless steel tanks respectively. It is evident in both cases that increasing the number of tanks, thus making them smaller and less massive, results into an higher demisability. For both materials, the increase in the Liquid Mass Fraction is considerable and can actually make the configuration completely demisable. In the stainless steel case it is possible to pass from configurations where none of the solutions have a demisability higher than 0.9 to having almost 40% of the solutions above this limit. This is a considerable increase in demisability. An interesting aspects is also represented by the fact that for the configurations with lower amount of tanks (1 to 3), just changing the material from stainless steel to aluminium alloy would not make the configuration completely demisable on average. Therefore, a further increasing in the number of tanks would be needed, or an integration of other design-for-demise solutions in order to increase the demisability.

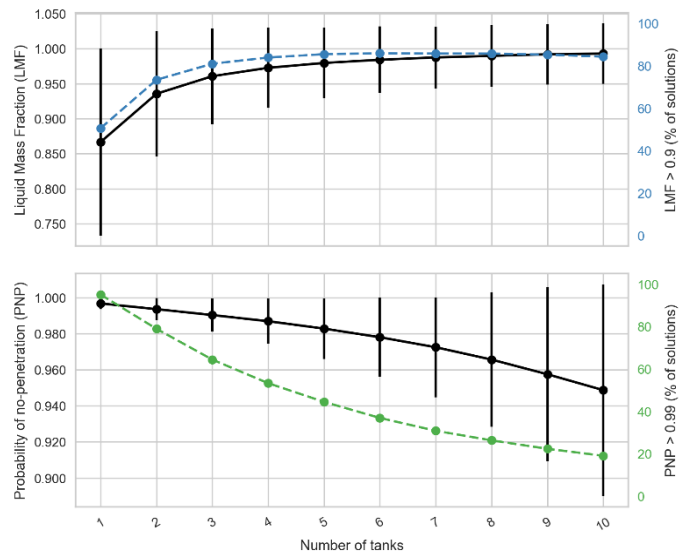

**Fig. 15: Variation of the average value and standard deviation of the *LMF* and *PNP* indices for variable dimensions of aluminium alloy componenents (black lines). Percentage of solutions over a 0.9 *LMF* index (blue line) and over a 0.99 *PNP* index (green lines).**

For what concerns the survivability, it is evident that increasing the number of components reduces the probability of no-penetration, at the cost, however, of increasing the configuration weight. Despite a higher number of elements produces smaller components, these components are also thinner and thus more vulnerable to the debris impact. In addition, more components, even smaller in dimensions, have in total a larger exposed area. For these reason we can observe a contrasting behaviour between the two indices when changing the dimensions. It is also important to observe the difference between the change in survivability produced for the aluminium alloy and for the stainless steel. In the first case, the reduction in the PNP index is about 5%, whereas in the second case is about 0.2%. There is thus one order of magnitude difference in the effects for the two materials. As such, changing the number of component for more vulnerable materials (such as aluminium alloys) is more influential on the overall probability of no-penetration of a configuration than for a more resistant material. Another aspect is the robustness of the solution. In the aluminium alloy case, the percentage of solution with a *PNP* index over 0.99 drops quickly with increasing number of tanks, down to 20% for a ten tanks configuration.

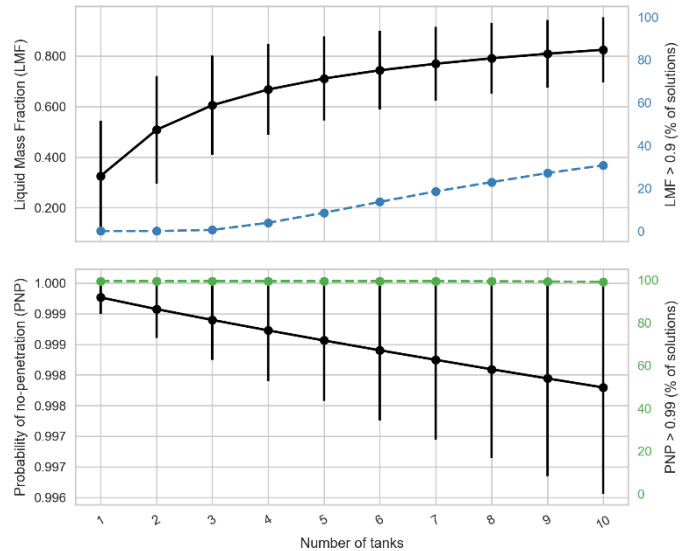

**Fig. 16: Variation of the average value and standard deviation of the *LMF* and *PNP* indices for variable dimensions of stainless steel componenents (black lines). Percentage of solutions over a 0.9 *LMF* index (blue line) and over a 0.99 *PNP* index (green lines).**

*4.4.4 Changing the component shape*

A change in the component shape can be reflected into changes into the demisability and the survivability of a component. The demisability is influenced because of the different area-to-mass ratio and the different shape factors. In the same way, also the survivability is influenced by a change in the area-to-mass ratio and in the thickness of the component. As the reference component is a tank, the two shape considered for the comparison here are a sphere and a cylinder. In Table 17 and Table 18, the results for the two shapes are presented for two material, the aluminium alloy 7075-T6 and the stainless steel AISI-316. For the simulation, the internal volume of the component was kept constant as well as the average cross-sectional area. The thickness varies according to Eq. (43) for the cylinder and Eq. (44) for the sphere.

$$t_{s,\min} = \frac{K_{sf} \cdot p_{\max} \cdot r}{2 \cdot \sigma_u} \quad (44)$$

Both the demisability and the survivability indices are higher for cylindrical tanks than for spherical ones. The *LMF* index is more significantly influenced by a change in the shape. For the aluminium case, the *LMF* increases by 0.237, and 0.339 for the stainless steel case. The change obtained is definitely not negligible. However, whereas the percentage of solutions with a LMF index above 0.9 significantly increase for the aluminium solutions, the same cannot be observed for the stainless steel solutions where no solution could achieve such a level of demisability. On the other end, the survivability is less influenced by the shape change, with a 0.015 variation in the *PNP* index for the aluminium alloy and a 0.001 variation for the stainless steel case.

**Table 17: Variation of the average value of the *LMF* and *PNP* indices with changing shape for aluminium alloy tanks. Percentage of solutions over a 0.9 *LMF* index and over a 0.99 *PNP* index.**

|                  | Sphere | Cylinder |
|------------------|--------|----------|
| **LMF (average)** | 0.709  | 0.946    |
| **PNP (average)** | 0.932  | 0.947    |
| **LMF > 0.9 (%)** | 26.4   | 75.8     |
| **PNP > 0.99 (%)**| 14.6   | 19.3     |

**Table 18: Variation of the average value of the *LMF* and *PNP* indices with changing shape for stainless steel tanks. Percentage of solutions over a 0.9 *LMF* index and over a 0.99 *PNP* index.**

|                  | Sphere | Cylinder |
|------------------|--------|----------|
| **LMF (average)** | 0.0002 | 0.339    |
| **PNP (average)** | 0.998  | 0.999    |
| **LMF > 0.9 (%)** | 0      | 0        |
| **PNP > 0.99 (%)**| 97.46  | 99.45    |

It can be concluded from the analysis that a change of the shape in a sensitive component such as spacecraft tanks can produce, on average, significantly more demisable solutions for low-melting point alloys such as aluminium alloys. Table 17 clearly shows that the average LMF index substantially increase and, more importantly, the percentage of solutions with a high value of the LMF index increases, thus ensuring a higher probability of having a fully demisable configuration. On the other end, it is also showed that the variation of the shape has little effect on the demisability of high-melting point alloys such as stainless steel. In fact, despite the average LMF index increases, still no solution can achieve a significant demisability. As such, changing the shape of high-melting point alloys can only be used as a complementary design-for-demise solution, with other, more effective, options to be considered as main way of action.

*4.4.5   Changing the component aspect ratio*

Another possible strategy to act on the demisability of a component is to modify its aspect ratio. To compare different solutions, the baseline tank of Table 11 is adopted. Starting from this geometry, the aspect ratio was varied so that the internal volume of the tank is kept constant. The range of aspect ratios taken into account goes from a minimum of 0.5 to a maximum of 2.0. In addition, the thickness of the component is kept constant and equal to 2 mm. Fig. 17 shows the effect of changing the aspect ratio for an aluminium alloy tank. The average variations result quite small, in particular if compared with the influence of design options such as changing the material or the dimensions of the component. In general, however, increasing the aspect ratio will slightly increase the demisability while at the same time slightly decreasing the survivability of a component as it is expected. Such design option can thus only be used as a complement to other, more effective, design solution as it cannot determine by itself a major variation of either of the two indices.

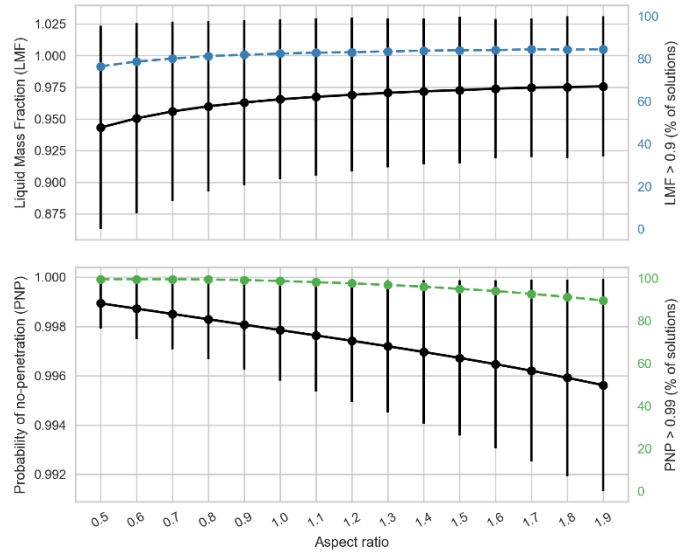

**Fig. 17: Variation of the average value and standard deviation of the *LMF* and *PNP* indices with changing aspect ratio for aluminium alloy componenents (black lines). Percentage of solutions over a 0.9 *LMF* index (blue line) and over a 0.99 *PNP* index (green lines).**

*4.5   Sensitivity to design-for-demise solutions*

Similarly, to the sensitivity analysis carried out in Section 4.3, and complementing the analysis of the previous paragraphs, the following section presents the sensitivity analysis of the demisability and survivability indices as a function of all the design-for-demise options taken into account (Section 4). Again, the Saltelli method (Section 4.1) is adopted with 2000 samples (equivalent to 24000 simulations) for both the demisability and the survivability. The ranges and values of the parameters considered in the sensitivity analysis are summarised in Table 19.

**Table 19: Ranges and values for the sensitivity analysis on the design-for-demise parameters.**

| Parameters | Ranges / Values | |
|---|---|---|
| Material | Al-6061-T6, Al-7075-T6, AISI-304, AISI-316, Graphite epoxy 1, Titanium 6Al4V | |
| Shape | Sphere, Cylinder | |
| Dimensions | 1 to 10 components | |
| Thickness | 0.0005 | 0.01 |
| Aspect ratio | 0.5 | 2.0 |

Fig. 18 shows the sensitivity of the demisability index to the design-for-demise options for a reference initial condition. A value close to 1 of the sensitivity index indicates that the considered variable significantly influences the demisability index of Eq. (30) as its variance with respect to the variable is large. A value close to zero instead is typical of parameter with a lower influence on the demisability index. The initial condition has an altitude of 80 km, a flight path angle of -1° and a relative velocity of 7.3 km/s. The conditions well represent an average uncontrolled re-entry from the LEO region with a break-up happening at an 80 km altitude. It is clearly observable that the most important parameter in determining the demisability of a component is the material. However, it is not always possible to change the material of a component, and other parameters can be changed in order to increase its demisability. Therefore, it is also important to consider the other parameters that play a role when adopting a design-for-demise approach.  Among the other parameters, the most important are the dimensions (width and radius) and the thickness. This could also be deduced by the results in Section 4.4.3 and 4.4.2 where the influence of changing the dimensions and the thickness was quantified for a reference component. Another interesting observation concerns the non-negligible difference between the first order effects and the total order effects, for the three main parameters (material, width/radius, and thickness). In fact, these three parameters greatly affect the ballistic coefficients of a re-entering component, which in turn affects the evolution of the re-entry trajectory. As such, not just their single contribution is important, but their combination.

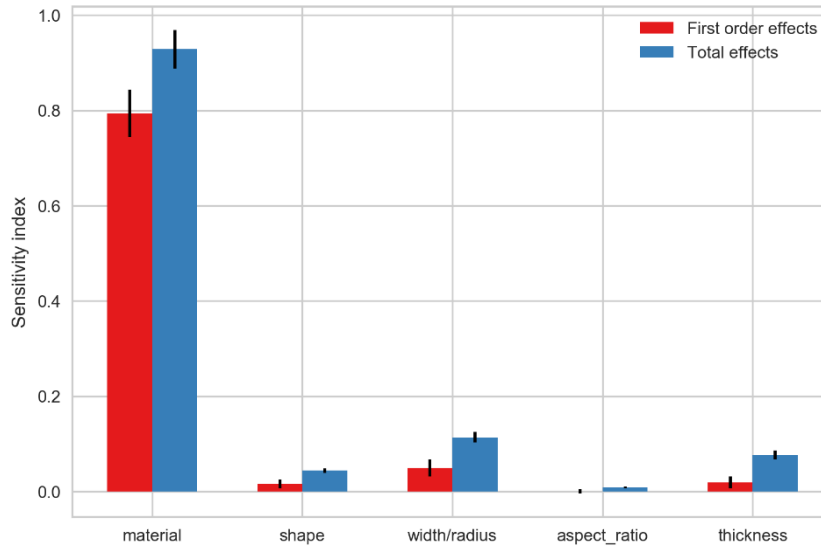

**Fig. 18: Sensitivity analysis of the demisability index with respect to the design-for-demise solutions.**

Fig. 19 shows an analogous sensitivity analysis for the survivability index. In this case, the most influential parameters are the thickness and the material, with a comparable sensitivity index. The dimensions have an average sensitivity on the survivability index, whereas the shape and the aspect ratio have a negligible influence on the output.

Another feature that can be observed is the difference between the first order and total order effects of the sensitivity. In the case of the demisability index, the first order effects are comparable to the total order effects, meaning that the effect of each parameter is not strictly coupled with the other parameters. On the other end, in the survivability case, the total order effects are clearly higher than the first order effects. Therefore, the contributions of the design parameters to the probability of no-penetration are coupled between each other. This could be observed in a quantitative way in the previous analysis. For example, the variation of the PNP index with the thickness (Fig. 13 and Fig. 14) has a clear difference in its variation when considering the aluminium alloys rather than the stainless steel, meaning that the effect of the change in thickness is also coupled to the change in the material.

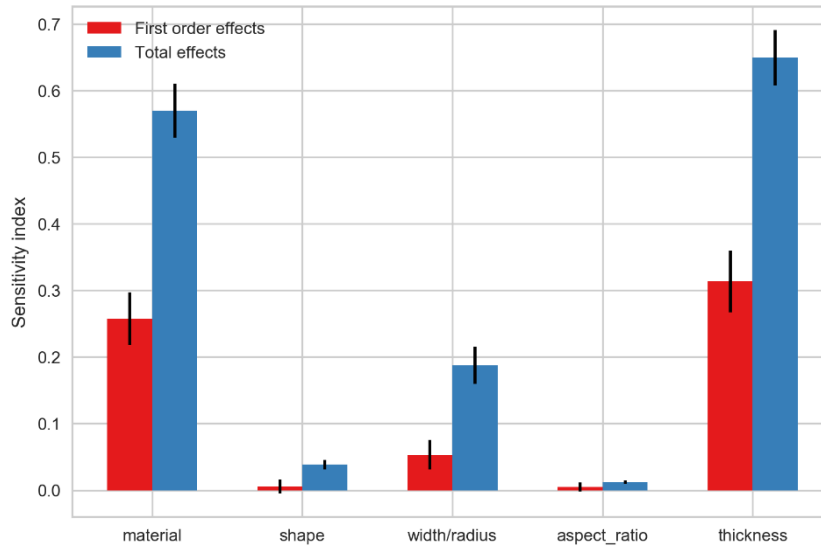

**Fig. 19: Sensitivity analysis of the survivability index with respect to the design-for-demise solutions.**

*4.6  Sensitivity analysis discussion*

The analysis presented in the previous sections shows the contribution of each design-for-demise option to the demisability and the survivability to spacecraft components. In addition, the sensitivity associated to each solution is evaluated for both the demisability and the survivability. Despite the results were obtained taking a specific component as a reference they are useful to determine the more important parameters to be considered when changing the design of a component following the design-for-demise principles. In fact, the reference component was selected to be representative of an actual component for which the design-for-demise principles are useful and can have an important impact with their application.

The most relevant design parameters are the material, the dimension, and the thickness of a component. As these parameters are important in influencing the demisability, they should be considered before the others when changing the design of a component. However, they are also the very influencing for the survivability. As such, a design has also to be verified against the survivability requirements when the design-for-demise options are implemented. This means that trade-off solutions should be searched where the different design options generate different levels of demisability and survivability. In a later stage, the most promising among these solutions are analysed in more detail and compared with the requirements and constraints of the mission design.

Alongside the identification of the most important design parameters, the quantitative influence of the single design options on the survivability and demisability indices was investigated. Many of the design options, except for some changes in material, affect the demisability and the survivability in a contrasting way that is while one of the two indices increases the other decreases. The different magnitude of this influence can also be observed, with the change of material and thickness being the most influential.

## 5  Demisability and survivability maps

The results of Section 4.5 show that the survivability and the demisability are mostly influenced by a subset of the design-for-demise options considered that are the material, the dimensions, and the thickness of the component. Following this consideration, a set of maps are presented for both the demisability and the survivability to better highlight the mutual dependency between these parameters. The maps show the variation of the demisability and survivability indices with the dimension and thickness for different shapes and materials. It is possible to generate such maps for the most common materials used in spacecraft design, for the main basic shapes and for a set of common initial conditions for the re-entry and most exploited orbits for the survivability. Such generated "booklet" of maps can be used as a useful tool to quickly assess the level of demisability and survivability that can be expected from a component, considering its dimension, thickness, and material, and to compare it with other design options. In fact, it is possible to locate on the maps the geometry considered and estimate its level of demisability and survivability. Then, moving on the maps is possible to consider the effect of changing the dimensions and the thickness. Jumping to another map, with the same geometry is instead possible to evaluate the effect of a change of material or shape.

*5.1 Demisability maps*

Examples of the aforementioned demisability maps are presented. Three commonly used materials are considered that are an aluminium alloy (Al-7075-T6), a stainless steel (AISI-316), and a titanium alloy (Ti-6Al-V), and three different shapes are taken into account (Sphere, Cylinder, Box). In addition, a set of six different orbit (Table 20) was used to present the difference between the maps.

**Table 20: Initial re-entry conditions examined.**

| Orbit | $h_0$ (km) | $V_0$ (km/s) | $fpa_0$ (°) | $lon_0$ (°) | $lat_0$ (°) | $head_0$ (°) |
|---|---|---|---|---|---|---|
| 1 | 100 | 7.3 | 0 | 0 | 0 | 45 |
| 2 | 100 | 7.3 | -1.5 | 0 | 0 | 45 |
| 3 | 100 | 7.3 | -3 | 0 | 0 | 45 |
| 4 | 80 | 7.1 | -1 | 0 | 0 | 45 |
| 5 | 80 | 7.1 | -2 | 0 | 0 | 45 |
| 6 | 80 | 7.1 | -3 | 0 | 0 | 45 |

Fig. 20 and Fig. 21 show how the demisability index (*LMF*) varies when changing the radius and the thickness of a spherical component. The map on the left correspond to the orbit case 3 of Table 20, and the map on the right to the orbit case 6. The radius ranges from 50 to 2500 mm while the thickness varies from 0.5 to 500 mm. The dark grey area in the bottom right corner of the plots represents a region of non-physical combinations of radius and thickness, i.e., the thickness is greater than the radius.

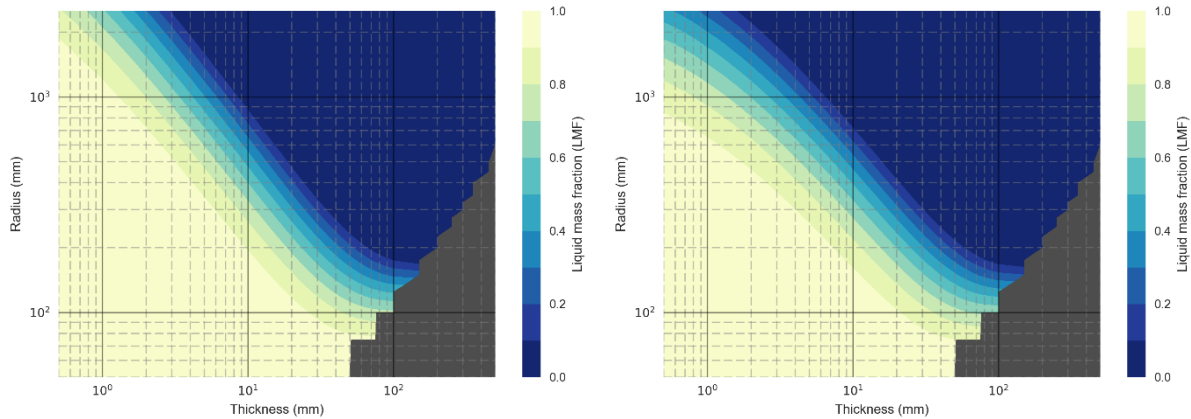

**Fig. 20: Demisability contour maps. Aluminium Sphere. Case 3 on the left and case 6 on the right.**

It is possible to observe from Fig. 20 that quite a big portion of the maps correspond to very demisable geometries, with a *LMF* index greater than 0.9. A regular pattern is also recognisable as the demisability reduces as going towards the upper right portion of the map. The two plots of Fig. 20 differ from each other for the initial altitude and velocity. In the plot on the right, given the lower initial altitude and velocity, the demisability reduces.

A considerable difference can be seen between the plot representing the aluminium sphere (Fig. 20) and the one of the stainless steel sphere (Fig. 21). The latter has an evidently smaller demisable region due to the higher melting temperature and heat capacity of the stainless steel with respect to the aluminium alloy, but also because of the larger mass of a steel sphere having the same dimension of an aluminium sphere. Again, the plots show a comparison between the initial conditions given by the orbit case 3 and 6 of Table 20. As expected, the plot on the right shows a smaller demisable are, shifted downwards of about a gridline in the logarithmic grid of the plot.

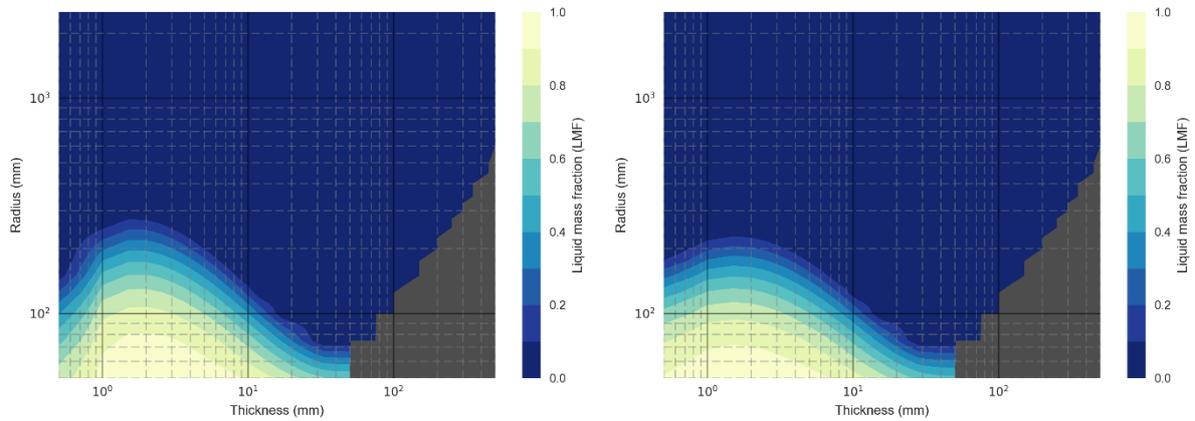

**Fig. 21: Demisability contour map. Stainless Steel Sphere. Case 3 on the left and case 6 on the right.**

Fig. 22 shows a comparison between the materials examined, with a simultaneous representation of the contours of the *LMF*. For sake of clarity, only three contours for each material are represented, showing the combination of radius and thickness where a 30%, 60%, and 90% of the object mass demises during the re-entry. The difference between the three materials is clear from the graph, with titanium being by far the less demisable of the materials, only granting a partial demise even for the smallest components. Follows the stainless steel and the aluminium alloy with increasing demisability.

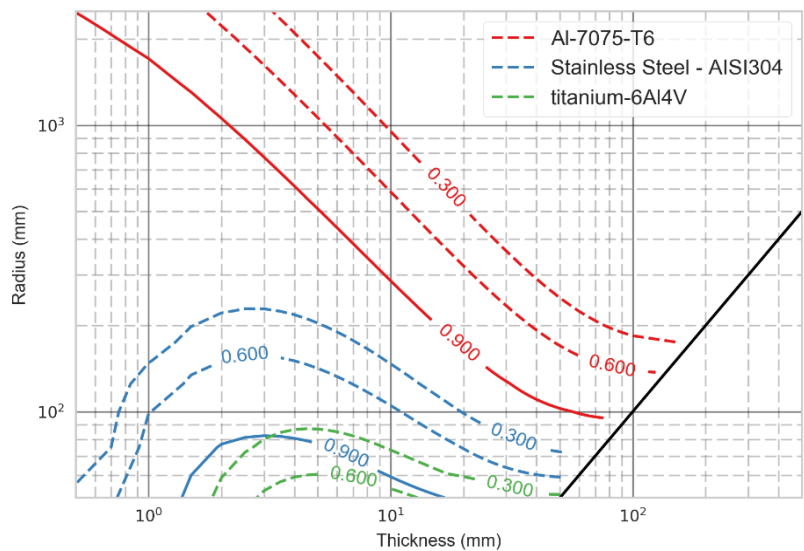

**Fig. 22: Comparison between the Liquid Mass Fraction of spheres made of aluminium 7075-T6, stainless steel, and titanium 6Al-4V for the orbit case 1**

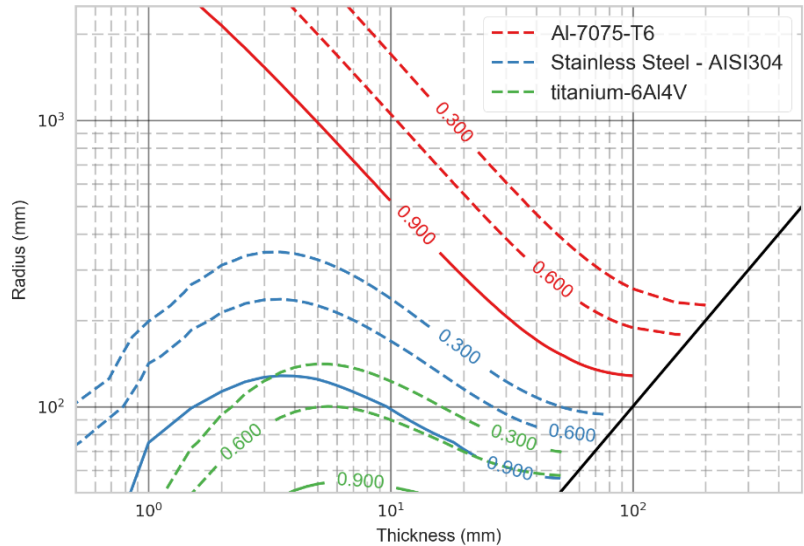

**Fig. 23: Comparison between the Liquid Mass Fraction of right cylinders made of aluminium 7075-T6, stainless steel, and titanium 6Al-4V for the orbit case 1**

In Fig. 23 and Fig. 24 equivalent diagrams for the other two shapes taken into account, the cylinder and the box are presented. Fig. 23 shows the same contours for a right cylinder, and Fig. 24 represents the contours for a cubic shape (all the sides have the same length). As it was mentioned at the beginning of the section, these types of maps can be readily used to estimate the demisability of a solution based on its dimensions, thickness, material, and shape. The maps can be either generated for different initial conditions, for example a range of initial velocities and break-up altitudes or a reference trajectory could be selected end evaluate the different options with respect to the results obtained with it. In this latter case, only one set of map has to be generated as function of the initial conditions. This last scenario is the one adopted at the ESA Clean Space office to compare different design-for-demise options.

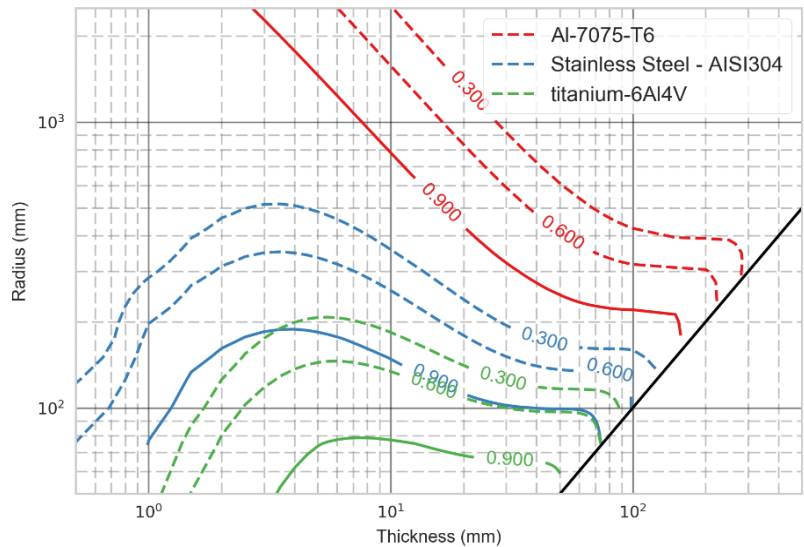

**Fig. 24: Comparison between the Liquid Mass Fraction of cubes made of aluminium 7075-T6, stainless steel, and titanium 6Al-4V for the orbit case 1.**

*5.2  Survivability Maps*

Analogous maps have been generated for the survivability of components as a function of their dimensions, thickness, material, and geometry. In addition, three different orbits have been considered

(Table 21), to take into account the variation in inclination and altitude for orbits in the sun-synchronous region that is one of the most exploited altitude range.

**Table 21: Characteristics of the examined orbits.**

| Characteristic | Orbit 1 | Orbit 2 | Orbit 3 |
|---|---|---|---|
| Altitude | 800 km | 700 km | 700 km |
| Eccentricity | 0.001 | 0.001 | 0.001 |
| Inclination | 98° | 30° | 60° |
| Mission time | 1 yr | 1 yr | 1 yr |

The maps of Fig. 25 and Fig. 26 represent the penetration probability variation with changing dimensions and thickness for a cubic structure made of aluminium alloy and stainless steel respectively. The orbital conditions used for the simulation are represented by the Orbit 2 case in Fig. 25.

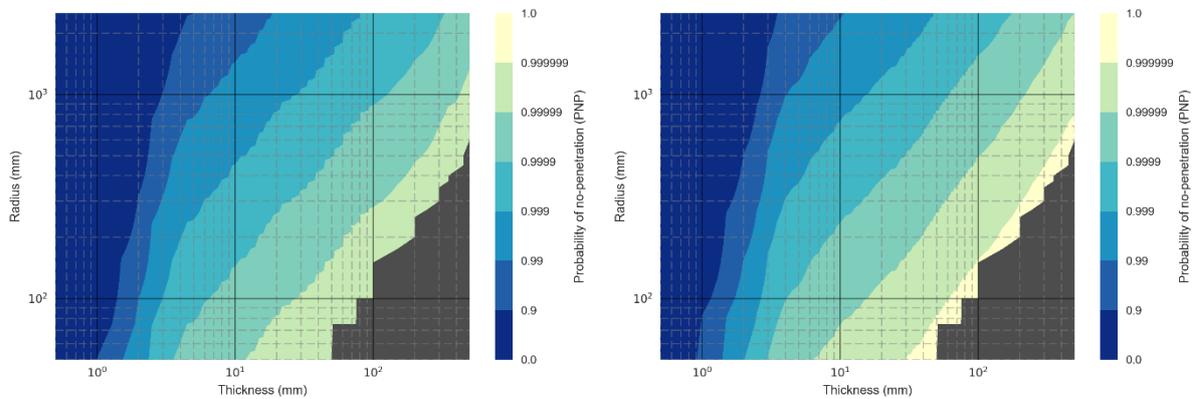

**Fig. 25: Survivability contour map. Aluminium 7075-T6 Cube. Orbit 1 case on the left, Orbit 3 case on the right.**

The plots have the same basic structure of the demisability contour maps with logarithmic scale on the axes, the same ranges of thickness and side length, and the non-physical zone in the bottom right corner of the plot. The contours of the map on the other hand are not linear anymore but they are represented in logarithmic scale ranging from 1 to 0, one being a 100% probability of no-penetration that is the structure basically provides full shielding.

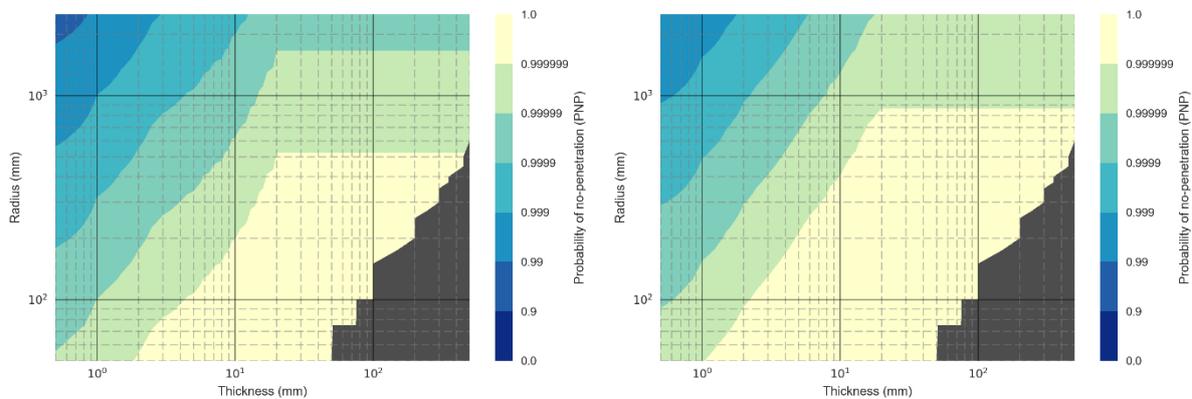

**Fig. 26: Survivability contour map - Stainless Steel Cube.**

It is possible to observe from both Fig. 25 and Fig. 26 that the behaviour for the survivability is the expected one, with the probability of no-penetration decreasing with decreasing thickness and dimension. It is possible to observe in Fig. 26 a flattening of the contour lines after a certain value of the side length, corresponding to a constant probability of no-penetration even with varying thickness. Examining at the Eq. (39) and considering a constant side length, the value of the area at risk remains constant and so does the mission duration when the mission is specified. Consequently, the only element changing is the critical flux whose value depends on the size of the critical diameter and on the shape of the particle flux distribution as a function of the diameter. Now, as the thickness increases, the critical diameter becomes bigger and can exceed the upper limit specified for the MASTER-2009 distributions. If this is the case, the critical flux is the one associated with the upper limit and is thus constant for all the thicknesses leading to a critical diameter greater than the limit one.

Considering the plots of Fig. 25, the difference between two operational orbits can be observed. The plot on the left represent an 800 km altitude orbit inclined by 98 degrees, whereas the plot on the left a 700 km orbit with a 60 degrees inclination. The right plot can be seen as shifted towards the upper left corner with respect to the left plot. Consequently, the areas with a lower probability of no-penetration are reduced in dimension, meaning that the orbit case 3 is less dangerous than the orbit case 1 for the same configuration. Comparing Fig. 25 and Fig. 26 the differences and the similarities between the maps of the two different materials (aluminium alloy and stainless steel) can be observed. It is clear that stainless steel structures are more resistant than aluminium structures having the same geometry by at least two orders of magnitude. The two graph, however, have a similar behaviour with almost linear contour lines that shifts from the bottom right corner to the upper left and vice versa according to the resistance of the material. The same happens when changing the orbit characteristics. This is a considerably different behaviour from the demisability where the shape of the maps was clearly influenced by the type of material. It is important to know and consider such trends when comparing different solutions and changing the characteristics of the components. Similarly to the demisability maps, the presented survivability maps.

A comprehensive map showing a comparison between the materials considered can be observed in Fig. 27. For the sake of clarity, only two contours for each material have been represented, showing the combination of side length and thickness where a 0.9 and a 0.999 *PNP* is achieved for the considered orbit and mission duration. Only the 0.999 line is present for the stainless steel case, meaning that such an option provides a very high shielding capability even at very low values of thickness. These kind of maps, alongside the previously presented demisability maps, can be generated for the different geometrical shapes considered and for the most exploited orbital regions, such as the sun-synchronous region (between 600 and 900 km altitude) and the constellation region (between 1200 and 1500 km of altitude). Such set of survivability maps can be used to quickly compare preliminary design by just looking at the position of the component inside the maps.

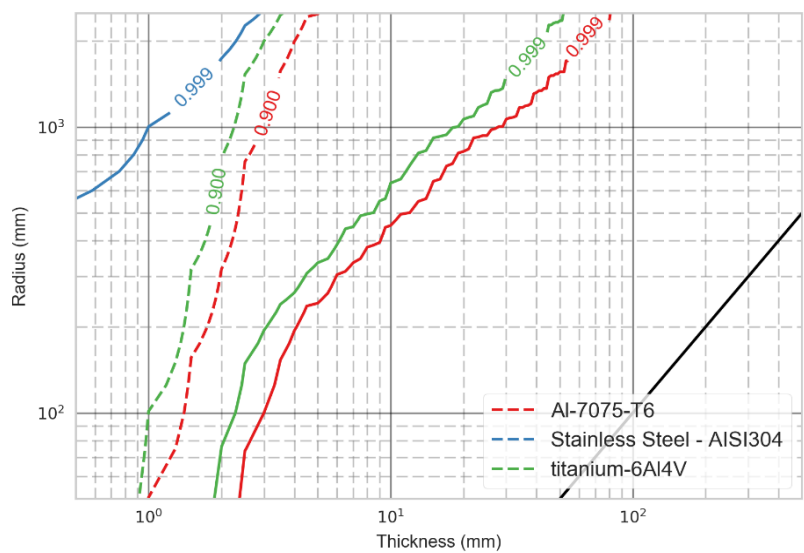

**Fig. 27: Comparison between the probability of no-penetration of cubes made of aluminium 7075-T6, stainless steel, and titanium 6Al-4V for the orbit case 2.**

# 6   Discussion and conclusions

Two models were developed to analyse the demise of re-entering spacecraft structures and components, and to assess the debris impact survivability during on-orbit operation. A demisability and survivability criteria were defined to assess the level of demisability and survivability of a specific design. The demisability criterion takes into account the amount of mass lost during the re-entry process and the impact energy threshold of 15 J. The survivability criterion relies on the definition of probability of-no penetration.

Using a reference component, the sensitivity of the demisability and survivability criteria as a function of the initial conditions to the models was studied. From the results, it is possible to observe that the demisability of components is mostly influenced by the initial break-up altitude and the velocity at break-up. A lower influence is represented by the flight path angle, whereas the longitude, latitude, and heading angles have a negligible sensitivity index. For the survivability, the sensitivity was studied as a function of the operational orbit altitude and inclination, together with the mission lifetime. All these parameters proved to be influential in influencing the survivability index, with the operational altitude being the most dominant parameter.

Having determined the most influential input parameters to the survivability and demisability models, our attention shifted towards the actual design options that can be used in a preliminary design phase to improve the demisability of components and ensure the compliance of the missions with the casualty expectation regulations. The main objective of this part of the work is to compare the different design-for-demise options and at the same time quantify their influence on both the demisability and the survivability. To do so, the different options have been analysed singularly and their effect on the demisability and survivability indices quantified. As the two indices depend upon the initial conditions, the study was performed over a range of input parameters, which were defined in the previous part of the work. As such, the behaviour of the demisability and survivability indices for each design option was assessed in terms of average and standard deviation. Concluding this section of the work a sensitivity analysis varying all the design-for-demise parameters was carried out for both the demisability and survivability index. The sensitivity results clearly showed that the most influential design parameter for the demisability is the material selection, whereas for the survivability the thickness is the most influential. In addition, while for the demisability the other parameters have a clear lower effect than the material, for the survivability, the material is similar in influence to the thickness and the dimensions have a considerable contribution. The results of the sensitivity analysis shows that, when considering the changing of a design using the design-for-demise principle, the most important parameter is clearly the material, followed by the thickness and the dimensions of the component. At the same time, these parameters have a clear influence on the survivability. It is thus important to consider both this aspects when implementing the design-for-demise options. This is especially true given the competing behaviour that many of the parameters exhibit with respect to the survivability and the demisability.

Finally, the paper presented some examples of demisability and survivability maps where the most influential design parameters were varied. These kind of maps can be used easily and effectively for a quick assessment and comparison of components design options. It is in fact possible to generate a set of maps covering the most common initial conditions for the re-entry, the most exploited orbital regions for the survivability, and do it for the different shapes, and a baseline set of materials. Once the maps are available, it is possible to locate on them the component considered and have a fast evaluation of both its demisability and survivability. In the same way, it is possible to compare different solutions and already select the most promising that will undergo a more detailed analysis in later stages of the mission design.

The paper has analysed the most common design-for-demise options and has assessed their influence on the demisability and the survivability of spacecraft components. Such influence has proved to be conflicting in nature for many of the parameters and the variation of the demisability and survivability when considering different design options has been quantified for each design-for-demise solutions separately. However, not all the design-for-demise options could be analysed given the simplified nature of the models used, which could change the relative sensitivity between the different design options. In addition, the computation of the survivability currently neglects the effect of shielding on internal components, which is a factor that could affect the relative weight of the sensitivity index of the survivability parameters.

Future work will focus first on the development of the demisability and survivability models. A three-level hierarchical structure of the models will be devised, with the parent spacecraft representing the first level, the internal components representing the second level, and a third level that can be used to model sub-components. In this way, a more realistic representation of a satellite configuration can be achieved. In addition, the possibility to specify the type of shielding for the main spacecraft structure will be added, including Whipple shield and honeycomb sandwich panel options. The survivability of internal component can in fact be significantly influenced by the type of shielding used for the main structure. Moreover, the possibility to consider the early detachment (before the main break-up of the spacecraft) of the external panels of the main structure will be included. It will also be possible to attach internal components to such panels in order to study the effect of an early detachment on the demisability of the internal components. The two improved models will then be used in a more integrated fashion in order to analyse preliminary spacecraft configurations with respect to the demisability and survivability. This will allow the study of the effects of design-for-demise options on both the demisability and the survivability since the early stages of the mission design. It will also allow taking into account the remaining design-for-demise options, which were not considered in this work. It will also allow the assessment of the influence of the shielding of internal components on the relative sensitivity of the survivability parameters.

**Acknowledgements**

This work was funded by EPSRC DTP/CDT through the grant EP/K503150/1. All data supporting this study are openly available from the University of Southampton repository at https://doi.org/10.5258/SOTON/D0280.
The authors would like to thank James Beck for his suggestions in the selection of heat flux correlations and shape factors.

**Appendix A – Material database data**

| Material | Density (kg/m$^3$) | Brinell hardness | Melting temp. (K) | Heat of fusion (J/kg) | Heat Capacity (J/kg/K) | Emissivity | Sound speed (m/s) | Yeld strength (MPa) |
|---|---|---|---|---|---|---|---|---|
| Al 6061 T6 | 2713 | 95 | 867 | 386116 | 896 | 0.141 | 5100 | 276 |
| Al 7075 T6 | 2787 | 150 | 830 | 376788 | 1012.35 | 0.141 | 5040 | 450 |
| Titanium 6Al4V | 4437 | 334 | 1943 | 393559 | 805.2 | 0.302 | 4987 | 880 |
| AISI304 | 7900 | 123 | 1700 | 286098 | 545.1 | 0.35 | 5790 | 215 |
| AISI316 | 8026.85 | 149 | 1644 | 286098 | 460.6 | 0.35 | 5790 | 250 |
| Inconel 601 | 8057.29 | n/a | 1659 | 311664 | 632.9 | 0.122 | 5700 | 450 |
| Graphite epoxy 1 | 1570 | n/a | 700 | 1.60E+07 | 1100 | 0.86 | n/a | 498.5 |
| Graphite epoxy 2 | 1550.5 | n/a | 700 | 236 | 879 | 0.9 | n/a | 498.5 |

**Table A1: Material database**